\documentclass[preprint,showpacs,preprintnumbers,amsmath,amssymb,natbib]{revtex4}

\usepackage{graphicx}
\usepackage{epsfig}		
\usepackage{dcolumn}
\usepackage{bm}
\usepackage{subfigure}
\def\0{\mbox{\tiny $0$}}
\def\1{\mbox{\tiny $1$}}
\def\2{\mbox{\tiny $2$}}
\def\3{\mbox{\tiny $3$}}
\def\4{\mbox{\tiny $4$}}
\def\5{\mbox{\tiny $5$}}
\def\6{\mbox{\tiny $6$}}
\def\7{\mbox{\tiny $7$}}
\def\8{\mbox{\tiny $8$}}
\def\9{\mbox{\tiny $9$}}

\def\f14{\mbox{\tiny $\frac{1}{4}$}}

\begin{document}

\title{Probing phase-space noncommutativity through quantum mechanics and thermodynamics of free particles and quantum rotors}

\author{Catarina Bastos\footnote{E-mail: catarina.bastos@ist.utl.pt}}
\affiliation{Instituto de Plasmas e Fus\~ao Nuclear, Instituto Superior T\'ecnico, Universidade de Lisboa, Avenida Rovisco Pais 1, 1049-001 Lisboa, Portugal}

\author{Alex E. Bernardini and Jonas F. G. Santos \footnote{E-mail: alexeb@ufscar.br, jonas@df.ufscar.br}} 
\affiliation{Departamento de F\'isica, Universidade Federal de S\~ao Carlos, PO Box 676, 13565-905, S\~ao Carlos, SP, Brasil.}


\date{\today}

\begin{abstract}
Novel quantization properties related to the state vectors and the energy spectrum of a two-dimensional system of free particles are obtained in the framework of noncommutative (NC) quantum mechanics (QM) supported by the Weyl-Wigner formalism.
Besides reproducing the magnetic field aspect of a Zeeman-like effect, the momentum space NC parameter introduces mutual information properties quantified by the quantum purity related to the relevant coordinates of the corresponding Hilbert space.
Supported by the QM in the phase-space, the thermodynamic limit is obtained, and the results are extended to three-dimensional systems.
The noncommutativity imprints on the thermodynamic variables related to free particles are identified and, after introducing some suitable constraints to fix an axial symmetry, the analysis is extended to two- and- three dimensional quantum rotor systems, for which the quantization aspects and the deviation from standard QM results are verified.
\end{abstract}

\pacs{03.65.-w, 03.67.-a, }
\keywords{phase space noncommutativity - thermodynamic limit - quantum rotors}
\date{\today}
\maketitle

\section{Introduction}

The intersection between quantum and classical descriptions of free particle systems \cite{Zurek} and the construction of relative thermodynamic variables associated to their statistical interpretation \cite{temp} have been historically discussed in the context of theoretical physics, since from superconductivity modeling \cite{superc} up to black hole physics \cite{BH}, and more recently, in the investigation of manifestly emergent phenomena \cite{Emerg} and of AdS/CFT holographic theories \cite{holog}.
Imprints on the thermodynamic variables related to free particle systems, including that for which an axial symmetry induces a quantum rotor-like description, can be indeed relevant in identifying tiny deviations from standard predictions of quantum mechanics (QM) on all the above-mentioned scenarios.

In the context of noncommutative (NC) deviations from standard QM predictions, the noncommutativity affects the quantum mechanical partition function and related thermodynamic variables displayed through the coupling of NC degrees of freedom to a thermal reservoir \cite{Bernardini13A}.
Noncommutativity in the coordinate space has firstly appeared as a suitable tool for the regularization of quantum field theories \cite{Snyder47}.
Considering that the space-time structure might be changed at the Planck energy scale, the noncommutativity have also been supposed to play a relevant role at quantum theories for gravitation.
In the same scope, NC geometry is also found in the context of string theory/M-theory \cite{Connes,Douglas,Seiberg}.

Likewise, the NC formulation of QM has supported the understanding of coupling and decoherence  aspects exhibited by two-dimensional quantum oscillators \cite{Rosenbaum,Bernardini13A}, the invertigation of the quantum Hall effect \cite{Prange,Belissard}, a clarifying analysis of electrons coupled to a magnetic field in the Landau level problem \cite{Nekrasov01}, and the suitable calculation of quantum corrections for the ultra-cold neutron energy spectrum in the scope of the gravitational quantum well problem \cite{06A,07A}.
Quantum effects driven by the phase-space NC QM have also been quantified \cite{Catarina} and related to missing information and quantum correlation issues \cite{Bernardini13A,Bernardini13B}.
Suitable features also include presumed violations of the Robertson-Schr\"odinger uncertainty relation \cite{Catarina001,Stein,Bernardini13C}, the regularizing features in minisuperspace quantum cosmology models \cite{Bastos00}, and also in black-hole physics \cite{Bastos004,Bastos005}.

Quantization properties related to the state vectors and the energy spectrum of a two- and three- dimensional ($2D$ and $3D$) systems have been preliminarily investigated in the context of NC QM \cite{Catarina,Catarina001,Catarina002,OB01}.
As prescribed for harmonic oscillators in QM, noncommutativity also introduces conceptual subtleties in describing the thermodynamics of quantum systems \cite{Bernardini13A}.
Hence, the relevant point examined in this contribution concerns the thermodynamic limit related to the phase-space NC extension of QM, given that NC modifications on measurable thermodynamic variables can all be determined, namely the NC correspondence to the internal energy, $U$, the Boltzmann entropy, $S_k$, and the heat capacity, $C_v$.
To establish the connection to the thermodynamic limit, the effect of noncommutativity between momentum coordinates, $[p_i,p_j]\neq0$, is preliminarily scrutinized in the context of the propagation of free particles with translational and rotational degrees of freedom. 
The correspondent thermodynamic limits of $2D$ and $3D$ gases of quantum systems of translationally free particles and of quantum rotors are then quantified.

The outline of this manuscript is then as follows.
In section II, the Weyl-Wigner (WW) formalism \cite{Wigner} and its effectiveness on defining the Groenewold-Moyal {\em star}-product \cite{Groenewold,Moyal} are reported in order to describe the NC extension of the QM in the phase-space. This formalism is then applied to calculate the NC properties of free particles in $2D$.
Besides identifying some analogy with preliminary results on the {\em star}genvalue problem of the harmonic oscillator in the NC phase-space \cite{Bernardini13A}, the phase-space time-evolution that leads to the time-dependence of the Wigner function is obtained.
One also identifies how noncommutativity induces the quantum decoherence of a state vector that is obtained through a consistent quantum preparation procedure.
The possibility of observing noncommutativity through local decoherence effects for a system of $2D$ free particles is discussed in section III.
The internal energy, the Boltzmann entropy, and the heat capacity, which provide the relevant elements for quantifying the thermodynamic limit are obtained in section IV.
In particular, the inclusion of a third dimension is also considered in order to shown that some aspects of the axial symmetry of the problem allow for a factorization of the NC effects.
Finally, extensions to the investigation of the thermodynamic limit of gases of $2D$ and $3D$ quantum rotors are examined in detail.
Our conclusions are drawn in section V.

\section{The Wigner function in the NC framework for free particles}

In the WW formulation of QM, composite quantum systems can be described in terms of the density matrix, $\hat{\varrho} = |\chi \rangle \langle \chi |$, such that one can identify the corresponding Wigner function through the Weyl transform as \cite{Wigner,Ballentine}
\begin{equation}
\mathcal{W}(q, p) =  h^{-1} \varrho^{\mathcal{W}} = 
h^{-1}\int \hspace{-.15cm}dy\,\exp{\left[i \, p \,y/\hbar\right]}\,
\chi(q - y/2)\,\chi^{\ast}(q + y/2),
\end{equation}
which, via the density matrix interpretation of $\varrho^{\mathcal{W}}$, allows one to compute the expectation value of an observable $\hat{O}$ through 
\begin{equation}
\langle O \rangle = 
\int \hspace{-.15cm}\int \hspace{-.15cm} dq\,dp \,\mathcal{W}(q, p)\,{O^{\mathcal{W}}}(q, p).
\end{equation}
Given that $\chi(q)$ can be identified by
\begin{equation}
\chi(q) =  h^{-1/2}\int \hspace{-.15cm} dp\,\exp{\left[i \, p \,q/\hbar\right]}\,\varphi(p),
\end{equation}
straightforward manipulations result into probability distributions for $q$ and $p$ as
\begin{equation}
\int \hspace{-.15cm} dq\,\mathcal{W}(q, p) = \varphi^{\ast}(p)\,\varphi(p)
\quad\mbox{and}\quad
\int \hspace{-.15cm} dp \,\mathcal{W}(q, p)= \chi(q)^{\ast}\chi(q).
\label{stdist1}
\end{equation}

From the intrinsic properties of the above definitions, one notices that an expression for the Wigner function can also be given in terms of $\varphi(p)$ as
\begin{equation}
\mathcal{W}(q, p) = 
\int \hspace{-.15cm}dy\,\exp{\left[-i\, q \,y/\hbar\right]}\,
\varphi(p - y/2)\varphi^{\ast}(p + y/2).
\end{equation}
Relevant properties which are connected to results from the density matrix formalism can thus be obtained through the WW prescription for QM \cite{Hillery,Lee}.
For instance, regarding the quantum information issues, the definition of quantum purity is given in terms of
\begin{equation}
Tr[\hat{\varrho}^2] =  h^{-1}\int \hspace{-.15cm}\int \hspace{-.15cm} dq\,dp \,\mathcal{W}(q, p)^2,
\end{equation}
where, in this case, the multiplication by a normalization factor, $h^{-1}$, is consistent with the density matrix framework that sets $Tr[\hat{\varrho}^2] = Tr[\hat{\varrho}] = 1$ for pure states\cite{Ballentine,Case}.

An important example of the usefulness of the WW formalism \cite{Wigner} is related to the functional implementation of the commutative Heisenberg-Weyl algebra of the ordinary QM,
\begin{equation}
\left[ \hat{Q}_i,  \hat{Q}_j \right] = 0 , \hspace{0.5 cm} \left[ \hat{Q}_i,  \hat{P}_j \right]
= i \hbar \delta_{ij} , \hspace{0.5 cm} \left[ \hat{P}_i,  \hat{P}_j \right] = 0 ,
\hspace{0.5 cm} i,j= 1, ... ,d ~. 
\label{stNCeq}
\end{equation}
Once an extension to a $2d$-dimensional phase-space is assumed, $\{\mathbf{Q},\mathbf{P}\}$, the Weyl transform turns quantum operators, $O(\mathbf{Q},\mathbf{P}; t)$, into $c$-numbers written as \cite{Rosenbaum,Catarina,Bernardini13A}
\begin{equation}
O^\mathcal{W}(\mathbf{Q},\mathbf{P}; t) = \int\hspace{-.1cm}\int\hspace{-.1cm}d\mathbf{x}\,d\mathbf{y}
F(\mathbf{x},\mathbf{y}; t) \, \exp{\left[\frac{i}{\hbar}\left(\mathbf{x}\cdot\mathbf{P} +\mathbf{y}\cdot\mathbf{Q}\right)\right]},
\label{stWigWig}
\end{equation}
and such a $2d$-dimensional generalization scheme leads to
\begin{equation}
F(\mathbf{x},\mathbf{y}; t) = h^{-d}\, Tr_{\{\mathbf{Q},\mathbf{P}\}}
\left[O(\mathbf{Q},\mathbf{P}; t)\,\exp{\left[\frac{i}{\hbar}\left(\mathbf{x}\cdot\mathbf{P} +\mathbf{y}\cdot\mathbf{Q}\right)\right]}
\right].
\label{stWeyWey}
\end{equation}
The quantum algebra from Eq.~(\ref{stNCeq}) can be reproduced by means of the Groenewold-Moyal {\em star}-product \cite{Groenewold,Moyal} which is supported by the relation
\begin{equation}
h \star  g = \exp{\left[\frac{\kappa}{2}\epsilon_{ij}\partial_{r_i}\partial_{s_j}\right]}h(r)\,g(s)|_{r=s},
\end{equation}
in the space of commutative functions, where $\kappa$ is suitably chosen to bring up the NC information.
Therefore, the Weyl-Wigner-Groenewold-Moyal (WWGM) phase-space formalism sets that
\begin{equation}
\langle \chi |O_{1} ( {\mathbf Q}, {\mathbf P}; t) O_{2}
( {\mathbf Q}, {\mathbf P}; t) | \chi \rangle 
= \int\hspace{-.15cm} \int \hspace{-.15cm} 
d{\mathbf P}\, d{\mathbf Q} \,
\varrho^{\mathcal{W}}(\mathbf{Q},\mathbf{P}; t)\,
O^{\mathcal{W}}_{1} (\mathbf{Q},\mathbf{P}; t) \star  O^{\mathcal{W}}_{2}(\mathbf{Q},\mathbf{P}; t), 
\label{stwigner}
\end{equation}
with 
\begin{equation}
\varrho^{\mathcal{W}}(\mathbf{Q},\mathbf{P}; t) =h^{-d}
\int d{\mathbf z} \exp[\frac{i}{\hbar} {\bf z}\cdot{\mathbf P}]\langle
{\mathbf Q} -\frac{{\mathbf z}}{2} |\varrho |{\mathbf Q} + \frac{{\mathbf
z}}{2}\rangle. 
\label{stdist}
\end{equation}
In particular, one notices that $\varrho^{\mathcal{W}}(\mathbf{Q},\mathbf{P}; t)$ reproduces the same analytical structure of the Wigner quasi-probability distribution given by Eq.~(\ref{stWigWig}).
The Heisenberg dynamics in terms of quantum operators is denoted by
\begin{equation}
\dot{O}^\mathcal{W}( {\mathbf Q}, {\mathbf P}; t) = -\frac{i}{\hbar} \left[O^\mathcal{W}(0), H^\mathcal{W}( {\mathbf Q}, {\mathbf P}; t)\right]_{\star} = \frac{i}{\hbar} (H^{\mathcal{W}} \star O^{\mathcal{W}} - O^{\mathcal{W}}\star H^{\mathcal{W}}),
\label{stmoyal2} 
\end{equation}
for which the Moyal product is parameterized by a bidifferential {\em star}-operator implemented through the exponential representation given by
\begin{equation}
\star =  \exp \left[
\frac{i\hbar}{2} ( {\overleftarrow{\nabla}_{{\mathbf Q}}}\cdot
{\overrightarrow{\nabla}_{{\mathbf P}}} - {\overleftarrow{\nabla}_{{\mathbf
P}}}\cdot {\overrightarrow{\nabla}_{{\mathbf Q}}} ) \right],
\label{stmoyal} 
\end{equation}
through which, the integral representation can be written as
\begin{equation}
\begin{split}
    O^{\mathcal{W}}_{1} ( {\mathbf Q}, {\mathbf P}) \star O^{\mathcal{W}}_{2} ( {\mathbf Q}, {\mathbf P})  =
    (2\pi\hbar)^{-2d}\int\dots\int d{\mathbf P}^{\prime} \ d{\mathbf P}^{\prime\prime}
      d{\mathbf Q}^{\prime} \ d{\mathbf Q}^{\prime\prime}  O^{\mathcal{W}}_{1} ({\mathbf P}^{\prime}, {\mathbf Q}^{\prime})
       O^{\mathcal{W}}_{2} ({\mathbf P}^{\prime\prime}, {\mathbf Q}^{\prime\prime})\\
       \exp[-\frac{2i}{\hbar}({\mathbf P}\cdot({\mathbf Q}^{\prime} - {\mathbf Q}^{\prime\prime})
       + {\mathbf P}^{\prime}\cdot({\mathbf Q}^{\prime\prime} - {\mathbf Q}) +
        {\mathbf P}^{\prime\prime}\cdot({\mathbf Q} - {\mathbf Q}^{\prime}))]. \label{stint1}\\
\end{split}
\end{equation}
Finally, one obtains that
\begin{equation}
\int\int d{\mathbf P} \ d{\mathbf Q} \ H^{\mathcal{W}} ( {\mathbf Q}, {\mathbf P}) \star\varrho^{\mathcal{W}} ( {\mathbf Q}, {\mathbf P})
= \int\int d{\mathbf P} \ d{\mathbf Q} \ H^{\mathcal{W}} ( {\mathbf Q}, {\mathbf P})
 \varrho^{\mathcal{W}} ( {\mathbf Q}, {\mathbf P}) = E, \label{stint2}
\end{equation}
which is an essential tool for parameterizing {\em star}genvalue problems involving quantum operators supported by a deformed Heisenberg-Weyl NC algebra.

In the scope of our analysis, the Heisenberg-Weyl algebra is modified as to have commutation relations given by
\begin{equation}
\left[ \hat{q}_i,  \hat{q}_j \right] = i \theta_{ij} , \hspace{0.5 cm} \left[ \hat{q}_i,  \hat{p}_j \right] = i \hbar \delta_{ij} ,
\hspace{0.5 cm} \left[ \hat{p}_i,  \hat{p}_j \right] = i \eta_{ij} ,  \hspace{0.5 cm} i,j= 1, ... ,d,
\label{stEq31}
\end{equation}
where $\eta_{ij}$ and $\theta_{ij}$ correspond to invertible antisymmetric  $d \times d$ matrices with real constant elements, such that an auxiliary matrix is defined by
\begin{equation}
\Upsilon_{ij} \equiv \delta_{ij} + {1\over\hbar^2}  \theta_{ik} \eta_{kj}.
\label{stEq32}
\end{equation}
One notices that $\mathbf{\Upsilon}$ is equally invertible if $\theta_{ik}\eta_{kj} \neq -\hbar^2 \delta_{ij}$ \cite{Catarina} and, through the Seiberg-Witten (SW) map \cite{Seiberg}, which is a linear transformation, the deformed NC algebra can be mapped into the usual Heisenberg-Weyl algebra as in Eq.~(\ref{stNCeq}).
Parameterizing the SW map through real constant matrices, ${\bf A},{\bf B},{\bf C},{\bf D}$, it can be written in the form of
\begin{equation}
 \hat{q}_i = A_{ij}  \mathit{\hat{Q}}_j + B_{ij}  \hat{\Pi}_j, \hspace{1 cm}
 \hat{p}_i = C_{ij}  \mathit{\hat{Q}}_j + D_{ij}  \hat{\Pi}_j.
\label{stEq35}
\end{equation} 
The SW map from (\ref{stEq35}) constrained by the NC relations from Eq.~(\ref{stNCeq}) can be cast in the form of matrix equations \cite{Rosenbaum} given by
\begin{equation}
{\bf A} {\bf D}^T - {\bf B} {\bf C}^T = {\bf I}_{d \times d} \hspace{1 cm} {\bf A} {\bf B}^T - {\bf B} {\bf A}^T = {1\over\hbar} {\bf \Theta} \hspace{1 cm}
{\bf C} {\bf D}^T - {\bf D} {\bf C}^T = {1\over\hbar} {\bf N}~,
\label{stEq36}
\end{equation}
where the entries of ${\bf A}, {\bf B}, {\bf C}, {\bf D}, {\bf \Theta}, {\bf N}$ are respectively given by $A_{ij}, B_{ij}, C_{ij}, D_{ij} , \theta_{ij} , \eta_{ij}$ and the symbol $T$ does allusion to a matrix transposition operator.

\subsection{The NC phase-space results for free particles}

Given that re-discussing the mathematical grounds of the WWGM NC formalism and its general applicability to quantum systems, as performed in \cite{Catarina} and even in \cite{Bernardini13A}, is out of the scope of this work, one can consider that relating the noncommutativity properties with observable quantum phenomena will be sufficient to apply the formalism to the well-known free particle system.

Thus, in order to probe the quantum effects due to the NC phase-space properties, let us consider the Hamiltonian for a $2D$ free particle of mass $m$ on a NC plane,
\begin{equation}
\hat{H}_{FP}(\mathbf{q},\mathbf{p}) = \frac{\mathbf{p}^2}{2m},
\end{equation}
such that the coordinante and momentum noncommutativity is quantified through
\begin{equation}
\left[  \hat{q}_i,  \hat{q}_j \right] = i \theta \epsilon_{ij}\hspace{0.2 cm}, \hspace{0.2 cm} \left[  \hat{q}_i,  \hat{p}_j \right] = i \delta_{ij}\hbar 
, \hspace{0.2 cm} \left[  \hat{p}_i,  \hat{p}_j \right] = i \eta \epsilon_{ij}\hspace{0.2 cm}, \hspace{0.2 cm} i,j=1,2,
\end{equation}
with $\epsilon_{ij} = -\epsilon_{ij}$.
The map to the commutative operators is thus given in terms of a SW map,
\begin{equation}
 \hat{q}_i = \nu  \mathit{\hat{Q}}_i - {\theta\over2 \nu \hbar} \epsilon_{ij} {\Pi}_j \hspace{0.5 cm},\hspace{0.5 cm}  \hat{p}_i = \mu {\Pi}_i + {\eta\over 2 \mu \hbar} \epsilon_{ij}  \mathit{\hat{Q}}_j~,
\label{stSWmap}
\end{equation}
which is invertible for $\nu$ and $\mu$ parameters constrained by their intrinsic rapport
\begin{equation}
{\theta \eta \over 4 \hbar^2} = \nu \mu ( 1 - \nu \mu ),
\label{stconstraint}
\end{equation}
with $\theta\eta \lesssim \hbar^2$. 
In this case, the inverse map is set as
\begin{eqnarray}
\mathit{\hat{Q}}_i &=& \mu \left(1 - {\theta \eta\over\hbar^2} \right)^{- 1 / 2} \left( \hat{q}_i + {\theta\over 2 \nu \mu \hbar} \epsilon_{ij}  \hat{p}_j \right),\nonumber\\
\hat{\Pi}_i &=& \nu \left(1 - {\theta \eta\over\hbar^2} \right)^{-1 / 2} \left( \hat{p}_i-{\eta\over 2 \nu \mu \hbar} \epsilon_{ij}  \hat{q}_j \right),
\label{stSWinverse}
\end{eqnarray}
with the corresponding Jacobian,
\begin{equation}
\bigg{\vert}\bigg{\vert}{\partial (q,p)\over \partial (\mathit{Q}, \Pi)}\bigg{\vert}\bigg{\vert} = (\det {\mathbf \Omega})^{1/2}=1 - {\theta \eta\over\hbar^2}.
\end{equation}

In terms of the commutative variables, $\mathit{\hat{Q}}_i$ and $\hat{\Pi}_i$, the Hamiltonian 
is written as
\begin{equation}
H^{\mathcal{W}}_{FP}(\mbox{\bf \em Q},\mathbf{\Pi}) = \alpha^2\mbox{\bf \em Q}^2 +\beta^2\mathbf{\Pi}^2 + \gamma \sum_{i,j = 1}^2{\epsilon_{ij}\Pi_i \mathit{Q}_j},
\label{stHamilton}
\end{equation}
where vectors are denoted by $\mbox{\bf \em V} = (\mathit{V}_1,\mathit{V}_2)$, and
\begin{eqnarray}
{\alpha}^2 \equiv {\eta^2\over 8m \mu^2 \hbar^2},\quad
{\beta}^2 \equiv {\mu^2\over 2m},\quad
\mbox{and} \quad {\gamma} \equiv \frac{\eta}{2m\hbar},
\label{steq37}
\end{eqnarray}
with $2 \alpha\beta = \gamma$ and, as expected, the results do not depend on $\theta$ and $\nu$.

One notices that the commutative variables, $\mbox{\bf \em Q}$ and $\mathbf{\Pi}$, satisfy the Hamilton equations of motion as set through Eq.~(\ref{stmoyal2}) for the Hamiltonian (\ref{stHamilton}).
Thus one obtains the following set of coupled first-order differential equations,
\begin{eqnarray}
\dot{\Pi}_k &=& -\frac{i}{\hbar} \left[\Pi_k,\,H^{\mathcal{W}}_{FP}\right] = -2 \alpha^2\,\mathit{Q}_k - \gamma\,\varepsilon_{jk}\Pi_j,\nonumber\\
\dot{\mathit{Q}}_k &=& -\frac{i}{\hbar} \left[\mathit{Q}_k,\,H^{\mathcal{W}}_{FP}\right] =  ~~2 \beta^2\,\Pi_k - \gamma\,\varepsilon_{jk}\mathit{Q}_j,\qquad k,j = 1,2,
\label{steqs01}
\end{eqnarray}
which can be rewritten as two uncoupled third-order differential equations,
\begin{eqnarray}
\dddot{\Pi}_k + 4\gamma^2\dot{\Pi}_k = 0,\nonumber\\
\dddot{\mathit{Q}}_k + 4\gamma^2\dot{\mathit{Q}}_k = 0,
\label{steqs01}
\end{eqnarray}
from which one gets the solutions,
\begin{eqnarray}
\mathit{Q}_1(t)&=& \frac{1}{2}\left[\left(x + \frac{\pi_y}{m\gamma}\right) + \left(x - \frac{\pi_y}{m\gamma}\right)\cos(2\gamma t) + \left(y + \frac{\pi_x}{m\gamma}\right)\sin(2\gamma t)\right],\nonumber\\
\mathit{Q}_2(t)&=& \frac{1}{2}\left[\left(y - \frac{\pi_x}{m\gamma}\right) + \left(y + \frac{\pi_x}{m\gamma}\right)\cos(2\gamma t) - \left(x - \frac{\pi_y}{m\gamma}\right)\sin(2\gamma t)\right],\nonumber\\
\Pi_1(t)&=& \frac{1}{2}\left[\left(\pi_x - m\gamma\, y\right) + \left(\pi_x + m\gamma\, y\right)\cos(2\gamma t) + \left(\pi_y - m\gamma\, x\right)\sin(2\gamma t)\right],\nonumber\\
\Pi_2(t)&=&\frac{1}{2}\left[\left(\pi_y + m\gamma\, x\right) + \left(\pi_y - m\gamma\, x\right)\cos(2\gamma t) - \left(\pi_x + m\gamma\, y\right)\sin(2\gamma t)\right],
\label{stsolutions}
\end{eqnarray}
where $x = Q_1(0),\,y= Q_2(0),\,\pi_x = P_1(0),$ and $\pi_y = P_2(0)$ are the initial conditions, and $\gamma   = \eta/{2m\hbar}$ is the characteristic frequency of the system.
In this case, $\mbox{\bf \em Q}$ and $\mathbf{\Pi}$ may be interpreted as classical dynamical variables within the WWGM formalism.

\begin{figure}
\includegraphics[width= 15cm]{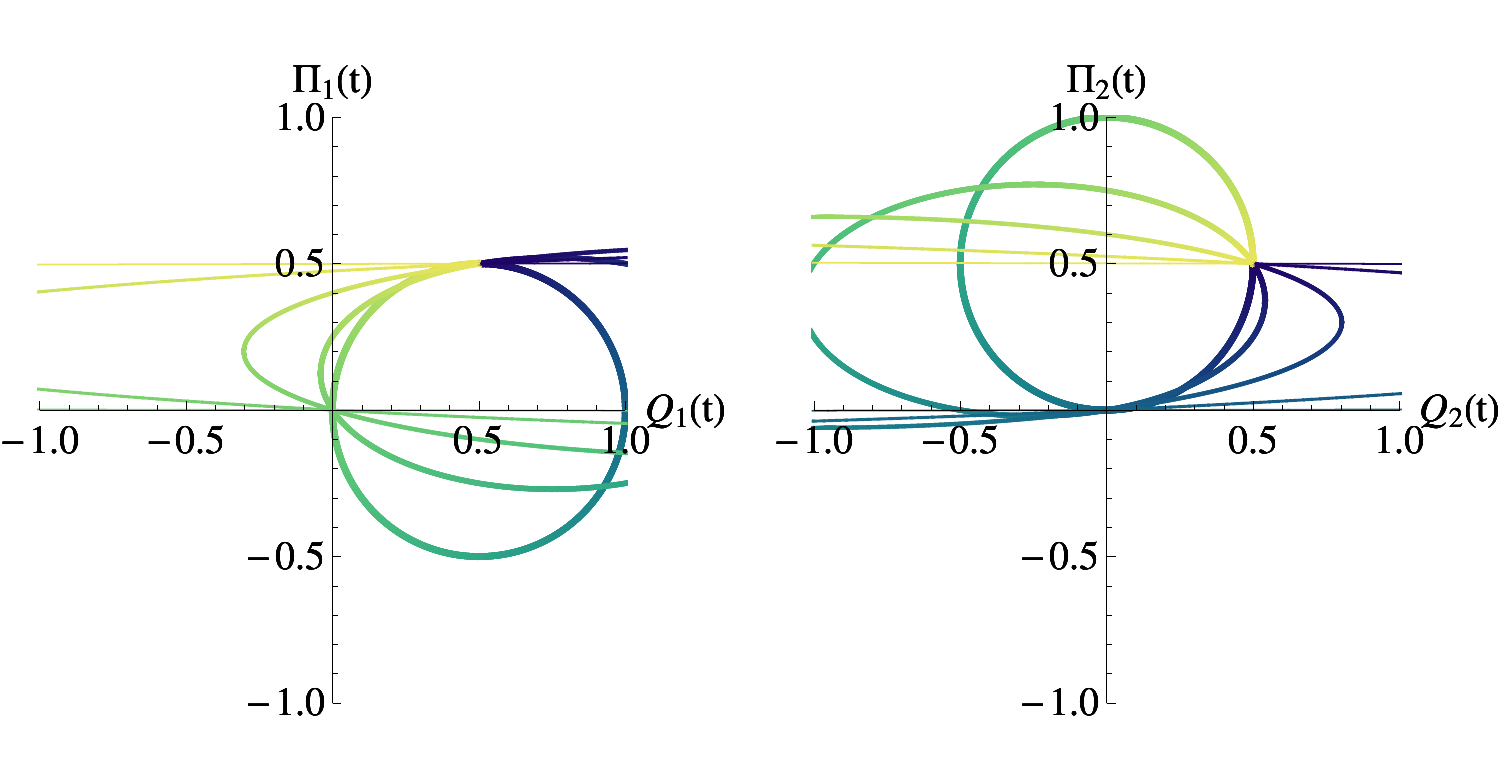}
\includegraphics[width= 1.cm]{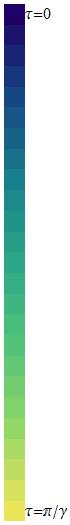}
\caption{\footnotesize  (Color online) Time evolution of the phase-space coordinates,$(\mathit{Q}_1(t),\Pi_1(t))$ and $(\mathit{Q}_2(t),\Pi_2(t))$, for the NC free particle system with the NC parameter, $\gamma$, being set equal to $1$ (thickest line), $1/2,\, 1/5,\, 1/20$, and $1/500$ (thinest line), implying a decreasing thickness.
One has used a {\em BlueGreenYellow} ({\em GrayLevel}) scale in order to denote the time scale, $\tau$, varying from $0$ (blue (dark gray)) to $\pi/\gamma$ (yellow (light gray)). By convenience, one has set $\hbar = m = 1$ and $\mathit{Q}_{1,2}(0)=\Pi_{1,2}(0)=0.5$. Notice that as $\gamma$ decreases, the phase-space trajectories approach to straight lines (standard QM free particle limit).}
\label{stPhase2}
\end{figure}

In particular, from Hamiltonian Eq.~(\ref{stHamilton}), one also identifies a characteristic Zeeman-like effect \cite{Ballentine} as it could be driven by a magnetic-like field, $B \sim \eta/\hbar q$, where $q$ is an ordinary electric charge.
Taking the limit of $\eta \rightarrow 0$, the magnetic field analogy disappears and Eqs.~(\ref{stsolutions}) are reduced to
\begin{eqnarray}
\mathit{Q}_1(t)&\sim& x + \frac{\pi_x}{m}t,\nonumber\\
\mathit{Q}_2(t)&\sim& y - \frac{\pi_y}{m}t,\nonumber\\
\Pi_1(t)&\sim& \pi_x ,\nonumber\\
\Pi_2(t)&\sim& \pi_y ,
\label{stsolutions2}
\end{eqnarray}
a typical free particle motion.
As it can be depicted in Fig.~\ref{stPhase2}, the time evolution of the phase-space coordinates, $(\mathit{Q}_1(t),\Pi_1(t))$ and $(\mathit{Q}_2(t),\Pi_2(t))$, are exhibited by elliptical curves for $\gamma \lesssim 1$, which can be extrapolated to straight lines in the limit where $\gamma$ vanishes.

The {\em stargen}functions for the Hamiltonian problem as set by Eq.~(\ref{stHamilton}) are determined through the {\em stargen}value equation,
\begin{equation}
H^{\mathcal{W}}_{FP} \star \varrho_n^{\mathcal{W}} (\mbox{\bf \em Q},\mathbf{\Pi}) = 
E_n\,\varrho^{\mathcal{W}}_n (\mbox{\bf \em Q},\mathbf{\Pi}),
\label{sthelp01}
\end{equation}
which, from the analysis developed in Ref.~\cite{Rosenbaum,Catarina,Bernardini13A}, and noticing that $2 \alpha\beta = \gamma$, results into
\begin{equation}
\varrho_{n}^\mathcal{W}(\mbox{\bf \em Q},\mathbf{\Pi}) = \mathcal{N}\frac{(-1)^n}{\pi \hbar}\mbox{exp}\left[-\Omega/\hbar \right]\, L_n^{0}\left(\Omega/\hbar\right),
\label{stLague01}
\end{equation}
which is written in terms of associated Laguerre polynomials, $L^0_n$, where $n$ is a non-negative integer and the energy spectrum is given by
\begin{equation}
E_{n} = \hbar \gamma(2 n + 1).
\label{stsener}
\end{equation}
In the above expression for $\varrho_{n}^\mathcal{W}(\mbox{\bf \em Q},\mathbf{\Pi})$,  $\Omega$ is a stationary variable given by
\begin{equation}
{\Omega}(t)= {\alpha\over\beta}\mbox{\bf \em Q}^2(t) + {\beta\over\alpha}\mathbf{\Pi}^2(t) + 2 \sum_{i,j = 1}^2{\left(\epsilon_{ij}\Pi_i(t) \mathit{Q}_j(t)\right)},
\end{equation}
i. e. by substituting Eqs.~(\ref{stsolutions}) into Eq.~(\ref{stLague01}), one indeed verifies that
\begin{equation}
\Omega(t) = \Omega(0) = {\alpha\over\beta}(x^2 + y^2) + {\beta\over\alpha}(\pi_x^2 +\pi_y^2) + 2(\pi_x \,y - \pi_y \, x),
\label{ststat}
\end{equation}
is a constant of motion.

\begin{figure}
\hspace{-1.5 cm}
\includegraphics[width= 17.8cm]{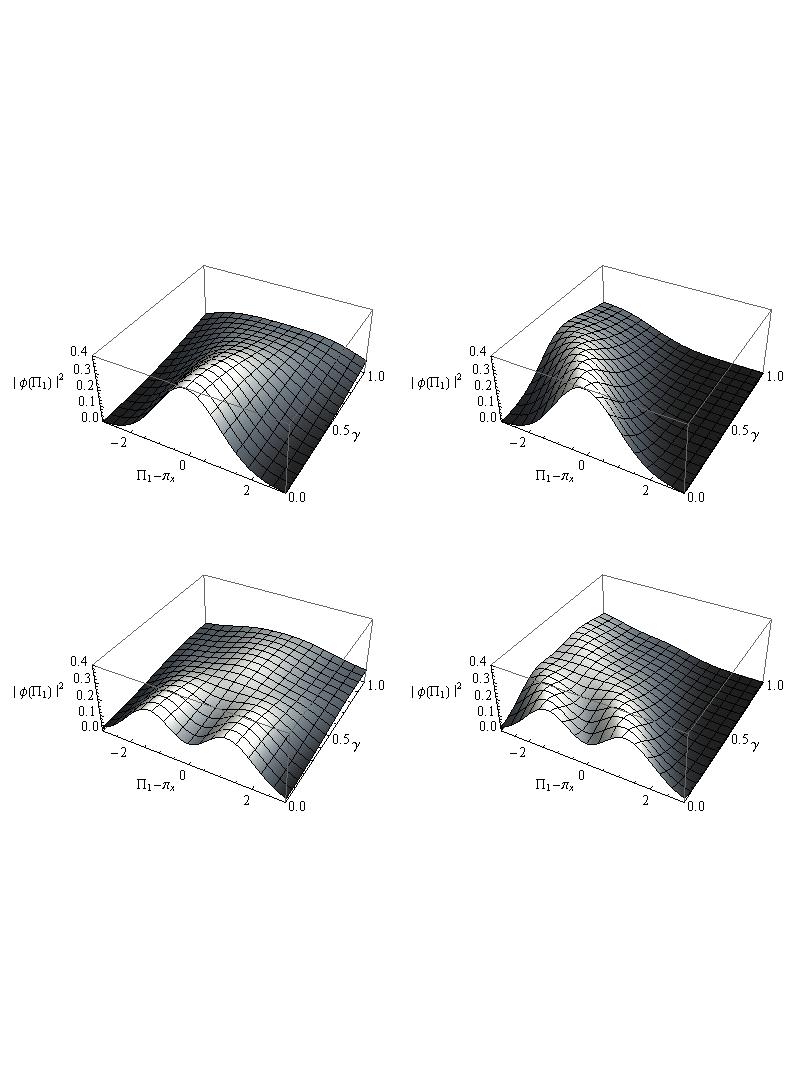}
\vspace{-6.0 cm}
\caption{\small Momentum distribution projection, $ \vert \varphi(\Pi_1;0)\vert^{2}$, onto $\Pi_1$ coordinate ($x$-direction) for the NC free particle system with associated NC quantum numbers $n = 0$ (first row) and $n = 1$ (second row). 
In order to distinguish the distortions introduced by the NC parameter, $\gamma$, one chooses two sets of $\mathit{Q}_2$ initial coordinates, i. e. $y = 0$  (first column) and $y = 4$ (second column), with $\pi_x$ arbitrary.
The results do not depend on $x$ and $\pi_y$.}
\label{stdist03}
\end{figure}

Given the free particle inherent features, one also notices that the normalization factor, $\mathcal{N}$, in Eq.~(\ref{stLague01}) has to be identified through an additional spatial localization constraint put over the free particle.
Noticing the constraint from Eq.~(\ref{ststat}), one integrates $\varrho_{n}^\mathcal{W}(\mbox{\bf \em Q},\mathbf{\Pi};t) \equiv \varrho_{n}^\mathcal{W}(\mbox{\bf \em Q},\mathbf{\Pi};0)$
over $\Pi_{1,2}$ from $-\infty$ to $+\infty$, and indeed obtains the probability distribution as an arbitrary constant.
Such a result demands for constraining $\mathit{Q}_{1,2}$ to finite intervals as to have $\mathit{Q}_{1,2} \in (-a,+a)$ in order to get a finite value for $\mathcal{N}$ (as performed for plane wave states).
Likewise, the integration over $\Pi_{2}$ and $\mathit{Q}_{2}$ then results into
\begin{eqnarray}
\int^{^{+a}}_{_{-a}}\hspace{-.5cm}d \mathit{Q}_2
\int^{^{+\infty}}_{_{-\infty}}\hspace{-.5cm}d \Pi_2\,
\varrho_{n}^\mathcal{W}(\mbox{\bf \em Q},\mathbf{\Pi};0) &=&
\tilde{\varrho}_{n}^{(1)}(\mathit{Q}_1,\Pi_1 ;0)
 \nonumber\\
&=&\tilde{\varrho}_{n}^{(1)}(\mathit{Q}_1,\Pi_1;0) \equiv (2 a)^{-1} \vert \varphi(\Pi_1;0)\vert^{2},
\end{eqnarray}
where, for the last step, one has noticed that $\tilde{\varrho}_{n}^{(1)}(\mathit{Q}_1,\Pi_1;0)$ effectively does not depend on $\mathit{Q}_1$, i. e. 
\begin{eqnarray}
\int^{^{+a}}_{_{-a}}\hspace{-.5cm}d \mathit{Q}_1\,
\tilde{\varrho}_{n}^{(1)}(\mathit{Q}_1,\Pi_1 ;0)
&=&2a\,\tilde{\varrho}_{n}^{(1)}(\mathit{Q}_1,\Pi_1 ;0) = \vert \varphi(\Pi_1;0)\vert^{2}.
\end{eqnarray}
Fig. ~\ref{stdist03} shows the stationary momentum distribution,  $\vert \varphi(\Pi_1;0)\vert^{2}$ for quantum numbers $n = 0$ and $1$, with $y = 0$ and $y = 4$, and $a$ arbitrarily set equal to $3$. The distributions do not depend on the parameters $x$ and $\pi_y$ and $\mathcal{N}$ is chosen as to give 
\begin{eqnarray}
\int^{^{+a}}_{_{-a}}\hspace{-.5cm}d \mathit{Q}_1
\int^{^{+\infty}}_{_{-\infty}}\hspace{-.5cm}d \Pi_1
\tilde{\varrho}_{n}^{(1)}(\mathit{Q}_1,\Pi_1;0) = 
\int^{^{+\infty}}_{_{-\infty}}\hspace{-.5cm}d \Pi_1
\vert \varphi(\Pi_1;0)\vert^{2} = 1.
\end{eqnarray}
The results show the distortion produced by the NC parameter $\gamma(\eta)$ on the momentum distribution profile.

\section{Quantum decoherence and missing information for Gaussian states in the momentum space}

Given that NC states like those ones from Eq.~(\ref{stLague01}) are stationary, obviously they are not suitable for identifying potential decoherence effects introduced by the NC QM.
To study how noncomutativity affects the evolution of vector states, it is more convenient to consider a Gaussian envelop for the momentum variables, such that eventual distortions introduced by the NC parameters can be suitably identified.

Generically, the WW formalism \cite{Case} sets that 
\begin{equation}
\varrho^{\mathcal{W}}_G(\mbox{\bf \em Q},\mathbf{\Pi};t) \equiv \varrho^{\mathcal{W}}_G(\tilde{x}(t),\tilde{y}(t),\tilde{\pi}_{x}(t),\tilde{\pi}_{y}(t);0),
\end{equation}
where $\tilde{x}(t),\,\tilde{y}(t),\, \tilde{\pi}_{x}(t)$ and $\tilde{\pi}_{y}(t)$ correspond to the inverse solutions of (\ref{stsolutions}) as to have, for instance, $\tilde{\pi}_{x,y}(t) \equiv \tilde{\pi}_{x,y}(\mathit{Q}_1,\Pi_1,\mathit{Q}_2,\Pi_2;t)$, that dictates the behavior of $\varrho^{\mathcal{W}}_G(\mbox{\bf \em Q},\mathbf{\Pi};t)$.
Hence, a normalized Wigner Gaussian envelop in the momentum space can thus be given by,
\begin{equation}
\varrho^{\mathcal{W}}_G(\mbox{\bf \em Q},\mathbf{\Pi};t) = \frac{1}{4 \pi a^2}\exp\left[-\left(\tilde{\pi}_x(t) - \pi_x)^2 + (\tilde{\pi}_y(t) - \pi_y)^2\right)\right],
\label{stgaussian1}
\end{equation}
where, again, the normalization factor follows the same constraints on $\mathit{Q}_{1,2}$, $\mathit{Q}_{1,2} \in (-a,+a)$.

Following the dynamics arising from $\tilde{\pi}_{x,y}(t) \equiv \tilde{\pi}_{x,y}(\mathit{Q}_1,\Pi_1,\mathit{Q}_2,\Pi_2;t)$ through the inverse of Eqs.~(\ref{stsolutions}), one can obtain the explicit time evolution for the state vector from Eq.~(\ref{stgaussian1}), $\varrho^{\mathcal{W}}_G(\mbox{\bf \em Q},\mathbf{\Pi};t)$.
Although it corresponds to a stationary state vector for standard QM parameters, for which $ \tilde{\pi}_{x,y}(t) \equiv \Pi_{1,2} = {\pi}_{x,y}$, it is noway stationary in case of NC dynamics.


One way to verify the effects of noncommutativity on the state vectors described by $\varrho^{\mathcal{W}}_G$ involves the calculation of the {\em reduced} Wigner functions (density matrices).
The time evolution of the {\em reduced} Wigner functions in the corresponding phase-space are identified by
\begin{eqnarray}
\tilde{\varrho}^{(1)}_G(\mathit{Q}_1,\Pi_1;t) &=&  Tr_{\{2\}}\left[\varrho^{\mathcal{W}}_G(\mbox{\bf \em Q},\mathbf{\Pi};t)\right]\nonumber\\
 &=& \int^{^{+a}}_{_{-a}}\hspace{-.5cm}d\mathit{Q}_2\int^{^{+\infty}}_{_{-\infty}}\hspace{-.5cm} d\Pi_2\,\, \varrho^{\mathcal{W}}_G(\mbox{\bf \em Q},\mathbf{\Pi};t),\\
\tilde{\varrho}^{(2)}_G(\mathit{Q}_2,\Pi_2;t) &=&  Tr_{\{1\}}\left[\varrho^{\mathcal{W}}_G(\mbox{\bf \em Q},\mathbf{\Pi};t)\right]\nonumber\\
 &=& \int^{^{+a}}_{_{-a}}\hspace{-.5cm}d\mathit{Q}_1\int^{^{+\infty}}_{_{-\infty}}\hspace{-.5cm} d\Pi_1\,\, \varrho^{\mathcal{W}}_G(\mbox{\bf \em Q},\mathbf{\Pi};t).
\label{sthelp05DD}
\end{eqnarray}
i. e. {\em integrating out} over $\mathit{Q}_{2,1}$ and $\Pi_{2,1}$ leads to the density matrix reduction of $\varrho^{\mathcal{W}}$ over these variables. The {\em reduced} Wigner function in the $\mathit{Q}_{1,2} - \Pi_{1,2}$ plane is obtained.

\begin{figure}
\includegraphics[width= 15.8cm]{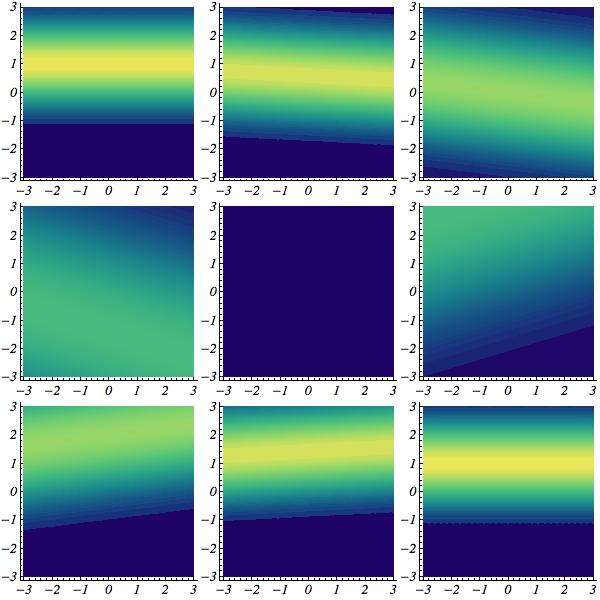}
\caption{\small  (Color online) Time evolution of the {\em reduced} Wigner function, $\tilde{\varrho}^{(1)}_G(\mathit{Q}_1,\Pi_1;t)$, for the NC free particle system corresponding to a state vector projected onto the $\mathit{Q}_1-\Pi_1$ plane.
At time $\tau = 0$ one has assumed that $\tilde{\varrho}^{(1)}_G(\mathit{Q}_1,\Pi_1;t)$ is centered at momentum $\pi_x = 1 (=-\pi_y)$, for an spatial coordinate arbitrarily given by $x = 1/\gamma$ (which is not relevant at $t = 0$).
One has considered time intervals such that $\tau = k \pi(8\gamma)^{-1}$, where $k$ corresponds to integer values from $0$ to $8$ (for plots from left to right and from up to down), and $m = 1$ (c. f. Eqs.~(\ref{stsolutions})).
The contour plot follows the {\em BlueGreenYellow} ({\em GrayLevel}) scale (from yellow (light gray) which corresponds to $1$, to blue (dark gray) which corresponds to $0$).}
\label{stMapa01}
\end{figure}

Considering that one looks over the time evolution of the state vectors along, for instance, the $x$ direction, the corresponding {\em reduced} Wigner functions, $\tilde{\varrho}^{(1)}_G(\mathit{Q}_1,\Pi_1;t)$ are depicted in Figs.~\ref{stMapa01}.
Notice that the NC parameter, $\gamma$, introduces a kind of periodical distortion of the departing stationary behavior such that the state vector, $\tilde{\varrho}^{(1)}_G(\mathit{Q}_1,\Pi_1;t)$, recovers its original wave pattern from $\tau = 0$ after a time $\tau =  \pi/\gamma$.
More specifically, $\gamma\neq 0$ modifies the expected behavior of the commutative free particle problem by introducing a spreading behavior to the localization character of the Wigner function, which also exhibits a phase-space rotating effect along the time evolution.
The quantum effects due to the NC parameter $\gamma$ are obtained by looking over time intervals as multiples of $\pi(8 \gamma)^{-1}$.

\begin{figure}
\includegraphics[width= 12cm]{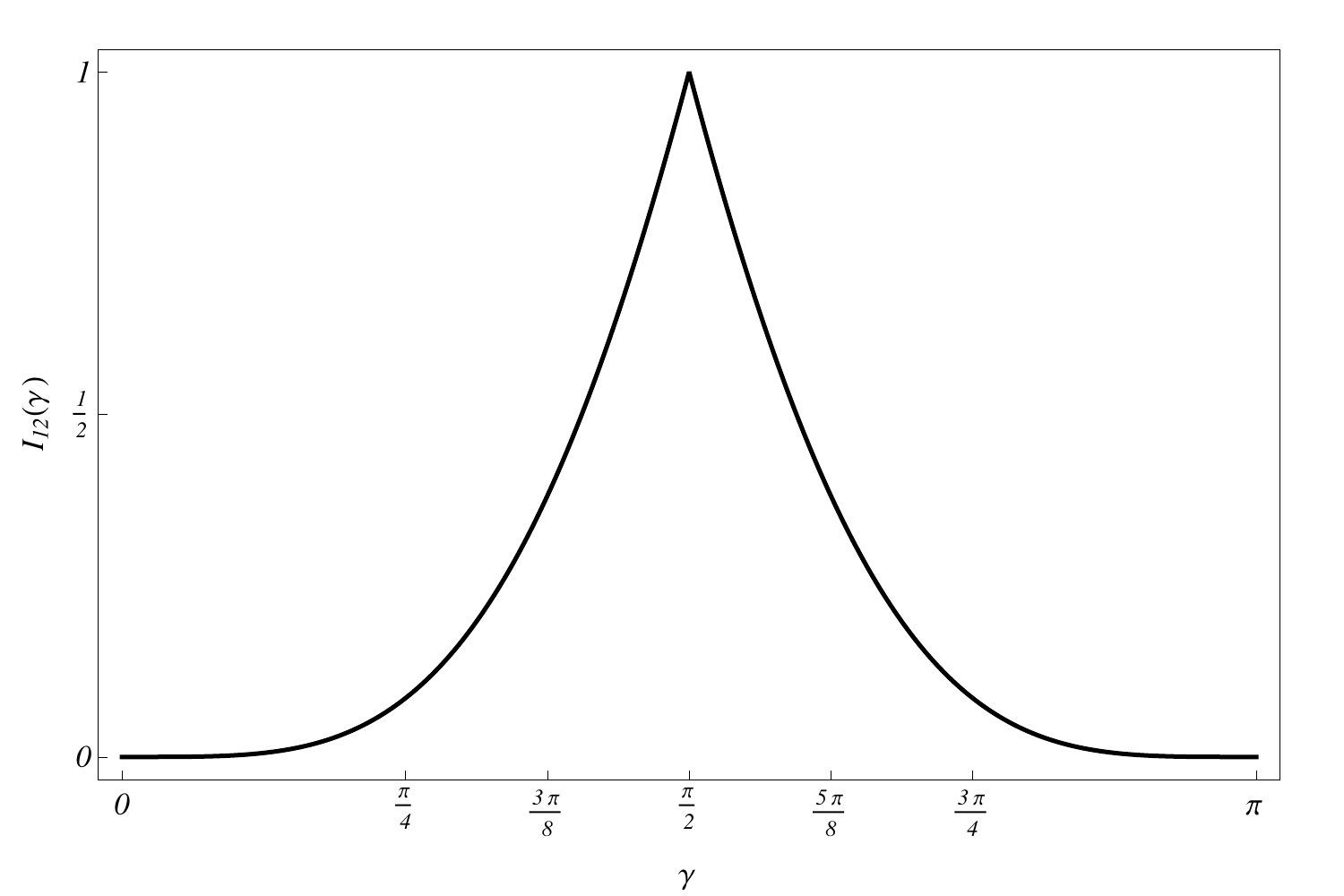}
\caption{\small (Color online) Mutual information, $I_{12}$, as function of $\gamma t$, for the NC free particle system described by the state vector $\varrho^{\mathcal{W}}_G$. As expected from Fig.~(\ref{stMapa01}), it quantifies the distortion of the stationary behavior and the oscillating character introduced by the NC parameter, $\gamma(\eta)$.}
\label{stMutual02}
\end{figure}

Moreover, to estimate the effects of the NC parameter, $\gamma$, on the time evolution of the quantum states like those from Eq.~(\ref{stgaussian1}), one may compute the linear entropy given in terms of quantum purity as
\begin{eqnarray}
S_1(t) &=& 1 - \frac{2\, a\sqrt{2\pi}}{\hbar}\, Tr_{\{1\}}\left[\left(Tr_{\{2\}}\left[\varrho^{\mathcal{W}}_G(\mbox{\bf \em Q},\mathbf{\Pi};t)\right]\right)^2\right] \nonumber\\
&=& 1 - \frac{2\, a\sqrt{2\pi}}{\hbar} Tr_{\{1\}}\left[\left(\tilde{\varrho}_G^{(1)}(\mathit{Q}_1,\Pi_1;t)\right)^2\right]\nonumber\\
&=&1 - |\cos(\gamma\, t)|,\nonumber\\
S_2(t) 
&=& 1 - \frac{2\, a\sqrt{2\pi}}{\hbar}\, Tr_{\{2\}}\left[\left(Tr_{\{1\}}\left[\varrho^{\mathcal{W}}_G(\mbox{\bf \em Q},\mathbf{\Pi};t)\right]\right)^2\right] \nonumber\\
&=& 1 - \frac{2\, a\sqrt{2\pi}}{\hbar}\, Tr_{\{2\}}\left[\left(\tilde{\varrho}_G^{(2)}(\mathit{Q}_2,\Pi_2;t)\right)^2\right]\nonumber\\
&=& 1 - |\cos(\gamma\, t)|, \nonumber\\
S_{12}(t) &=& 1 - \frac{8\, \pi\,a^2}{\hbar^2} Tr_{\{1\}}\left[Tr_{\{2\}}\left[\left(\varrho^{\mathcal{W}}_G(\mbox{\bf \em Q},\mathbf{\Pi};t)\right)^2\right]\right]\nonumber\\
&=&1 - \cos(\gamma\, t)^{2},
\label{stlinear}
\end{eqnarray}
through which a quantum correlation between internal subsystems quantified by the {\em mutual information} is straightforwardly computed as
\begin{equation}
I_{12}(t) = S_1(t) + S_2(t) - S_{12}(t) = I_{21}(t) = (1 -|\cos(\gamma\, t)|)^{2},
\end{equation}
which is depicted in Fig.~\ref{stMutual02}, and quantifies the mutual correlation between $x(\leftrightarrow 1)$ and $y(\leftrightarrow 2)$ states, in this case, exclusively due to the NC features.
With $\gamma = 0$, the mutual information, $I_{12}(t)$, as well as all the above defined entropies vanish, and $\varrho^{\mathcal{W}}_G(\mbox{\bf \em Q},\mathbf{\Pi};t)$  shall reproduce the product of two uncorrelated pure states, i. e. the state vector evolves like a stationary pure state \footnote{
In general, if two variables $x$ and $y$ are assumed to be uncorrelated, the quantum {\em mutual information} measures the discrepancy in the uncertainty resulting from such a presumable  erroneous hypothesis.}.
Therefore, through the above analysis, one identifies the momentum coordinate NC parameter, $\gamma(\eta)$, as the main agent in introducing a quantified quantum correlation between free particle Gaussian states in $2D$ momentum space.

\section{Partition function and thermodynamic variables for $2D$ and $3D$ NC free particles}

A gas of free particles in thermal contact with an environment at temperature $T$ corresponds to a {\em canonical} ensemble. 
For the case of $2D$ gases, the microstates occupied by the system obeying the NC properties of QM are described by $\varrho^{\mathcal{W}}_{n}$, where $E_{n}$ denotes the ({\em stargen})energy of the system in a given microstate (c. f. Eq.~(\ref{stsener})).
Given the coarse grained nature of $2D$ coordinate and momentum spaces, these microstates are regarded as a system of discrete quantum states labeled by the parameter $n$.
Formally, the partition function is obtained through the {\em trace} expressed in terms of the coherent state vector, $\varrho^{\mathcal{W}}_{n}$, as to have
\begin{eqnarray}
Z(\sigma) &=& 
\int^{^{+a}}_{_{-a}}\hspace{-.5cm}d \mathit{Q}_1
\int^{^{+a}}_{_{-a}}\hspace{-.5cm}d \mathit{Q}_2
\int^{^{+\infty}}_{_{-\infty}}\hspace{-.5cm}d \Pi_1
\int^{^{+\infty}}_{_{-\infty}}\hspace{-.5cm}d \Pi_2\,\,
Tr\bigg{[}\exp\left[-\frac{H^{\mathcal{W}}}{k_B\,T}\right] \,\varrho^{\mathcal{W}}_{n} \bigg{]}
\nonumber\\
&\equiv& \sum_{n = 0}^\infty \exp\left[-\frac{E_{n}}{k_B\,T}\right] 
=\exp[{-\sigma}]\sum_{n = 0}^\infty \exp\left[-2n\,\sigma\right]\nonumber\\
&=& \frac{1}{2}\frac{1}{\sinh{(\sigma)}},
\label{stpartition}
\end{eqnarray}
which is consistent with the above discussed normalization condition \cite{Bernardini13A} for Eq.~(\ref{stLague01}), and where we have introduced the parameter $\sigma = \hbar\gamma/k_B\,T$, from which the NC imprints would be evinced through the explicit dependence on $\gamma = \eta/2m\hbar$.

The relevant thermodynamic quantities for the $2D$ gas are straightforwardly computed from $Z(\sigma)$.
One has the internal energy given by
\begin{equation}
U(\sigma) = - \hbar \gamma \,\frac{\partial}{\partial \sigma} \ln{\left[Z(\sigma)\right]},
\label{stene}
\end{equation}
the Boltzmann entropy given by
\begin{equation}
S_k(\sigma) = -k_B \,\sigma^{\2}\frac{\partial}{\partial\sigma}\left[\frac{1}{\sigma}\ln{\left[Z(\sigma)\right]}\right],
\end{equation}
and the heat capacity given by
\begin{equation}
C_v = k_B \,\sigma^2\,\frac{\partial^2}{\partial \sigma^2}\ln{\left[Z(\sigma)\right]}.
\label{stcapa}
\end{equation}

The result from Eq.~(\ref{stpartition}) can be substituted into the above definitions as to give
\begin{equation}
U(\sigma) =  \hbar \gamma  \coth(\sigma),
\end{equation}
\begin{equation}
S_k(\sigma) = - k_B \left(\ln{\left[2\sinh{(\sigma)}\right]} -\sigma\,\coth(\sigma)\right),
\end{equation}
and
\begin{equation}
C_v = k_B\left(\frac{\sigma}{\sinh{(\sigma)}}\right)^{2}.
\end{equation}

Obviously $2D$ gases of free particles should exhibit NC properties mostly in the limit where $\sigma \rightarrow \infty$ such that, from the above defined thermodynamic variables, $U/\hbar\gamma$ tends to unity, in agreement with the quantum equipartition theorem, and and $C_v/k_B$ tends to zero.
This offers an additional possibility for measuring NC effects at very low temperatures.
The classical limit given by $\sigma \ll 1$ sets $U/\hbar\gamma\sim \sigma^{-1}$ and $C_v/k_B \sim 1$, as expected for a classical $2D$ gas of free particles.


\subsection{Inclusion of a third dimension}

One can pose the question about whether the NC effects exhibited by the $2D$ gas can be generalized to a $3D$ system of free particles.
In this context, it is helpful to consider the $3D$ extension of the previously discussed NC problem.
The Hamiltonian for the problem can then be written as
\begin{eqnarray}
\hat{H}_{3D}(\mathbf{q},\mathbf{p}) &=& \frac{\mathbf{p}^2}{2m} = \frac{1}{2m}\sum_{i=1}^{3} p_{i}^{2},
\end{eqnarray}
that leads to
\begin{eqnarray}
H^{\mathcal{W}}_{3D}(\mbox{\bf \em Q},\mathbf{\Pi}) &=& \frac{1}{2m}\left[\mathbf{\Pi} + \frac{1}{2\hbar}\left({\boldsymbol\eta}\times \mbox{\bf \em Q} \right)\right]^2 \nonumber\\
&=&\frac{\mathbf{\Pi}^{2}}{2m} + \frac{1}{2 m\hbar}{\boldsymbol\eta}\cdot\left(\mbox{\bf \em Q} \times \mathbf{\Pi} \right)
+\frac{1}{8m\hbar^2}\left[{\boldsymbol\eta}^2 \mbox{\bf \em Q} ^2 - ({\boldsymbol\eta}\cdot \mbox{\bf \em Q}) ^2\right],\label{stHamiltonian2A}
\end{eqnarray}
once one has considered a simplified version of the $3D$ SW map from Eq.~(\ref{stSWmap}), by setting $\mu = 1$ as to have
\begin{equation}
\hat{p}_i =  \hat{\Pi}_i + {1\over 2 \hbar} \epsilon_{ijk}  \eta_j\,\mathit{\hat{Q}}_k~,
\label{stSWmap2}
\end{equation}
which reflects the $3D$ NC relation between momenta,
\begin{equation}
\left[  \hat{p}_i,  \hat{p}_j \right] = i \epsilon_{ijk}\eta_{k}\hspace{0.2 cm}, \hspace{0.2 cm} i,j=1,2,3,
\end{equation}
where $\epsilon_{ijk}$ is the Levi-Civita tensor of rank three and ${\boldsymbol\eta} = (\eta_1,\eta_2,\eta_3)$.

Due to analogous arguments related to the axial symmetry exhibited by the QM $3D$ Zeeman effect, the Wigner distribution dependence on the third dimension can be factorized from its final form for the prescription from Eq.~(\ref{stHamiltonian2A}).
In particular, one chooses the appropriate representation for the quantum operators in a coordinate system where ${\boldsymbol\eta} = \eta_{3} \hat{z} \equiv \eta \hat{z}$ such that the $3D$ NC Hamiltonian from Eq~(\ref{stHamiltonian2A}) can be rewritten as (see, for instance, the resolution of problem 11.6, pag. 632, at Ref.~\cite{Ballentine})
\begin{equation}
H^{\mathcal{W}}_{3D}(\mbox{\bf \em Q},\mathbf{\Pi}) 
=\frac{\mathbf{\Pi}^{2}}{2m} + \frac{\eta}{2 m\hbar} L_z
+\frac{\eta^2}{8m\hbar^2}\sum_{j=1}^2\hat{Q}_j^2.
\label{stHNC$3D$modified}
\end{equation}

Ignoring any quantization property related to the factorized $z$ coordinate bounded by the interval $(-a,+a)$, the $3D$ version of the partition function can be written in terms of $Z(\sigma)$ from Eq.~(\ref{stpartition}) as
\begin{eqnarray}
Z^{3D}(\sigma) &=& Z(\sigma)\left(\frac{1}{h}
\int^{^{+a}}_{_{-a}}\hspace{-.5cm}dz
\int^{^{+\infty}}_{_{-\infty}}\hspace{-.5cm}d p_z
\exp\left[-\frac{p_z^{2}}{2m k_B\,T}\right]\right)
\nonumber\\
&=& 
\sqrt{\frac{2}{\pi}}\frac{a}{\hbar}
Z(\sigma)
\sqrt{m\,k_B T},
\label{stpartition$3D$}
\end{eqnarray}
which, besides the evident introduction of additional (not quantized) degrees of freedom, does not contribute to the scope of quantifying NC effects.

\subsection{Rotational degrees of freedom and the thermodynamics of NC quantum rotors}

Turning our attention to the degeneracies brought up by the angular momentum operators, namely $L^2_z$ and $L^2$, $2D$ and $3D$ Hamiltonians may shed some light on the study of noncommutativity, in particular, on the confront between the thermodynamic properties of $2D$ and $3D$ gases of quantum rotors.
One knows that, besides translational degrees of freedom, free particles with internal structures not only carry internal (electronic) degrees of freedom but also may have its kinetic behavior modified by rotations and vibrations.
Through a rigid rotator approach, the above prescription for NC free particles supports the discussion of NC quantum rotors in $2D$ and $3D$.

Introducing the rigid rotator axial constraint, $R^2 = \sum_{j=1}^2 Q_j^2$, and using $\gamma = \eta/2m\hbar$ into the Hamiltonian $H^{\mathcal{W}}_{2D}$, one has
\begin{equation}
H^{\mathcal{W}}_{2D}(L_z)
=\frac{L_z^2}{2 m R^2} + \gamma L_z
+\frac{1}{2}m R^2\gamma^2,
\label{stHL2}
\end{equation}
such that $H^{\mathcal{W}}_{2D}(L_z) \star \varrho^{\mathcal{W}}_{m_z} = E^{2D(R)}_{m_z}\varrho^{\mathcal{W}}_{m_z}$, for which the {\em stargen}energies are written as
\begin{equation}
E^{2D(R)}_{m_z}
=\hbar\gamma \left[\frac{m_z^2}{\lambda} + m_z + \frac{\lambda}{4} \right],
\label{stEL2}
\end{equation}
with $|m_z| = 0,\,1,\, \dots, \infty$, and where we have introduced a dimensionless parameter of {\em inertia}, 
\begin{equation}
\lambda = \frac{2 m R^2 \gamma}{\hbar} = \frac{R^2 \eta}{\hbar^2},
\end{equation}
that sets a magnitude of inertia relative to the NC inertia momentum, $I_{\eta} = m \hbar^2/\eta$, i. e. $I = mR^2 = \lambda I_{\eta}$.

The straightforward $3D$ generalization leads to $H^{\mathcal{W}}_{3D}$ given by
\begin{equation}
H^{\mathcal{W}}_{3D}(L^2,L_z)
=\frac{L^2}{2 m R^2} + \gamma L_z
+\frac{1}{2}m R^2\gamma^2,
\label{stHL3}
\end{equation}
such that $H^{\mathcal{W}}_{3D}(L_z) \star \varrho^{\mathcal{W}}_{m_z,\ell} = E^{3D(R)}_{m_z,\ell}\varrho^{\mathcal{W}}_{m_z,\ell}$, for which the {\em stargen}energies are written as 
\begin{equation}
E^{3D(R)}_{m_z,\ell}
=\hbar\gamma \left[\frac{\ell(\ell+1)}{\lambda} + m_z  + \frac{\lambda}{4} \right].
\label{stEL2}
\end{equation}
with $|m_z| = 0,\,1,\, \dots, \ell$ and $\ell = 0,\,1, \dots, \infty$.

The above results provide the tools for obtaining the respective partition function for ensembles of $2D$ and $3D$ quantum rotors.
A straightforward calculation gives
\begin{eqnarray}
Z^{2D(R)}_{NC}(\sigma) &=&
\sum_{m_z = -\infty}^\infty \exp\left[-\sigma\left(\frac{m_z^2}{\lambda} + m_z + \frac{\lambda}{4}\right)\right] 
\end{eqnarray}
for $2D$ gases, and
\begin{eqnarray}
Z^{3D(R)}_{NC}(\sigma) &=&
\sum_{\ell = 0}^\infty \sum_{m_z = -\ell}^\ell
\exp\left[-\sigma\left(\frac{\ell(\ell+1)}{\lambda} + m_z  + \frac{\lambda}{4}\right)\right] \nonumber\\
&=&\sum_{\ell = 0}^\infty 
\left[\cosh(\ell\sigma) + \coth(\sigma/2) \sinh(\ell\sigma) \right]
\exp\left[-\sigma\left(\frac{\ell(\ell+1)}{\lambda} + \frac{\lambda}{4}\right)\right] 
\end{eqnarray}
for $3D$ gases.
For both $2D$ and $3D$ cases, given the dependence of $\sigma$ and $\lambda$ on $\gamma$, the limits for which $\gamma$ tends to $0$ result into the standard QM partition functions, respectively,
\begin{eqnarray}
Z^{2D(R)}(\sigma) &=&
\sum_{m_z = -\infty}^\infty \exp\left[-\sigma\frac{m_z^2}{\lambda}\right],
\end{eqnarray}
and
\begin{eqnarray}
Z^{3D(R)}(\sigma) &=&
\sum_{\ell = 0}^\infty 
\left(2 \ell + 1\right)\exp\left[-\sigma\frac{\ell(\ell+1)}{\lambda}\right].
\end{eqnarray}
As previously introduced by Eqs.~(\ref{stene}-\ref{stcapa}), the relevant thermodynamic quantities computed from the above partition functions of NC quantum rotors can be straightforwardly obtained.

\begin{figure}
\includegraphics[width= 17cm]{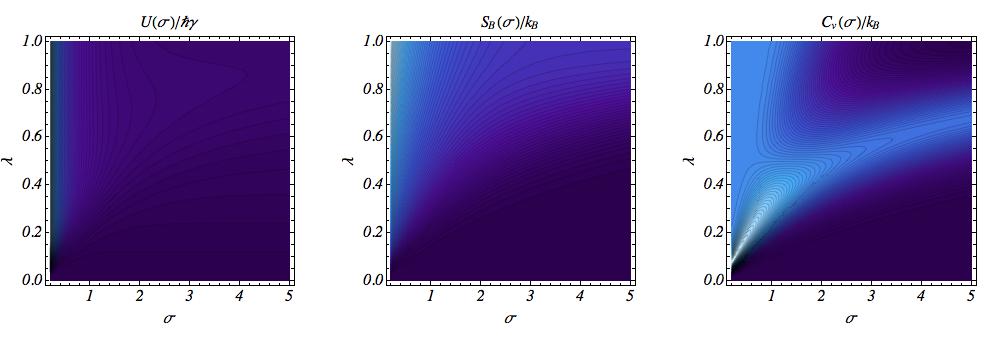}
\includegraphics[width= 17cm]{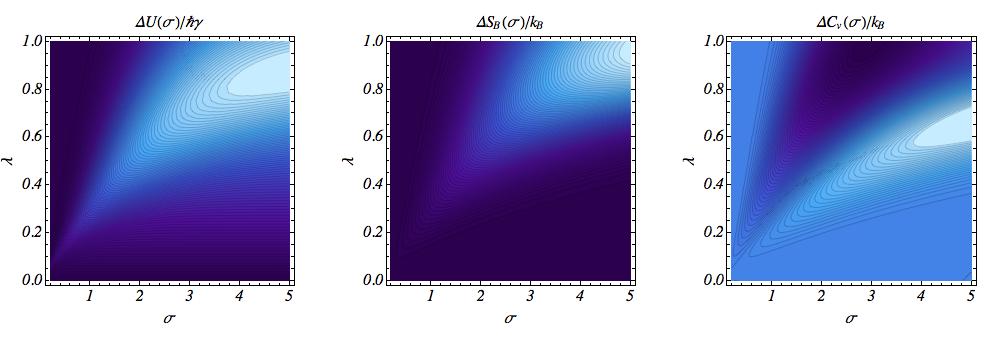}
\caption{(Color online) Internal energy, $U(\sigma)$, Boltzmann entropy, $S_k(\sigma)$, and the heat capacity, $C_v(\sigma)$, of a gas of $2D$ quantum rotors (first row), and the corresponding variations with respect to the analogous standard QM system (second row), as function of thermodynamic and inertia parameters, $\sigma$ and $\lambda$. 
A {\em Blue(Gray)Tone} scale is assumed with dark blue (gray) regions corresponding to zero, and light blue (gray) regions corresponding to unity.}
\label{stFig.1}
\end{figure}

\begin{figure}
\includegraphics[width= 17cm]{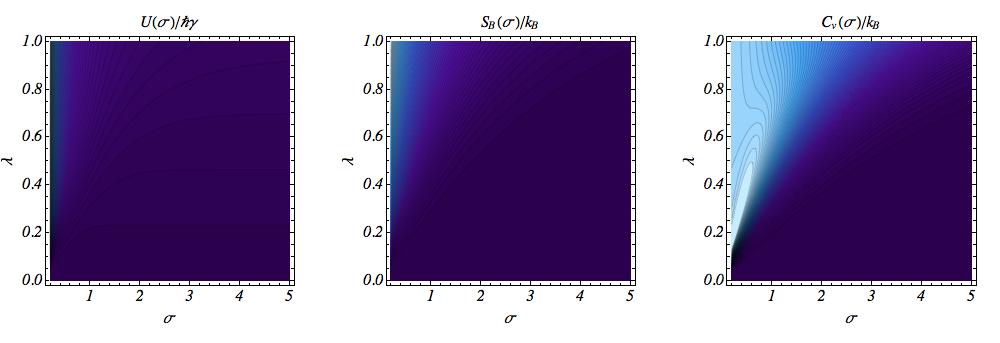}
\includegraphics[width= 17cm]{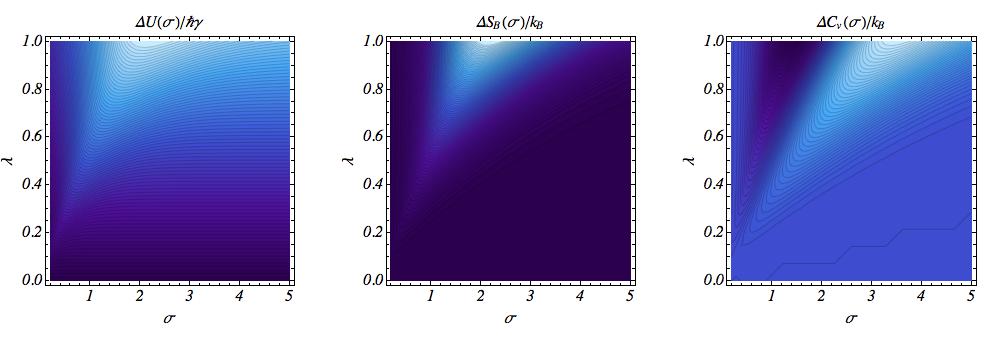}
\caption{Internal energy, $U(\sigma)$, Boltzmann entropy, $S_k(\sigma)$, and the heat capacity, $C_v(\sigma)$, of a gas of $3D$ quantum rotors (first row), and the corresponding variations with respect to the analogous standard QM system (second row), as function of thermodynamic and inertia parameters, $\sigma$ and $\lambda$. 
The color scheme is reproduced from Fig.~\ref{stFig.1}.}
\label{stFig.2}
\end{figure}

Figs.~\ref{stFig.1} and \ref{stFig.2} show the internal energy, $U$, the Boltzmann entropy, $S_k$, and the heat capacity, $C_v$, of NC quantum gases (first row), as function of the thermodynamic and inertia parameters, $\sigma$ and $\lambda$, and the corresponding $\Delta \mathcal{Q} = \mathcal{Q}^{(NC)} -\mathcal{Q}^{\tiny(Stand)}$ quantities (second row), with $\mathcal{Q} = U,\,S_k,\, C_v$, for comparing NC with standard QM results for $2D$ (Fig.~\ref{stFig.1}) and $3D$ (Fig.~\ref{stFig.2}) quantum gases. 
Qualitatively, two overall effects can be identified for $2D$ and $3D$ quantum gases.
First, the maximal deviation from standard QM results (dark region) occurs for finite and non-vanishing values of the thermodynamic parameter: $\sigma \sim 5$ for $2D$ gases and $\sigma \sim 2-3$ for $3D$ gases.
Second, increasing values of the inertia parameter, $\lambda$, up to certain arbitrary values evidently intensifies the NC deviation from standard QM results.
It has been an expected effect, given the role of $\lambda$ in the previously defined Hamiltonians.

Comparing $2D$ and $3D$ configurations, one also notices an additional effect: the inclusion of a third degree of freedom results into an amplification of the NC effects with respect to the temperature.
For $3D$ quantum rotors, the maximal deviation due to the NC modifications occurs at higher temperatures (lower values of $\sigma$) whether it is compared with those values of maximal deviation for $2D$ configurations.
Giving the general systematic difficulty of reaching smaller temperatures in the Nature, $3D$ configurations are more suitable for observing the NC effects.

\begin{figure}
\includegraphics[width= 9.3cm]{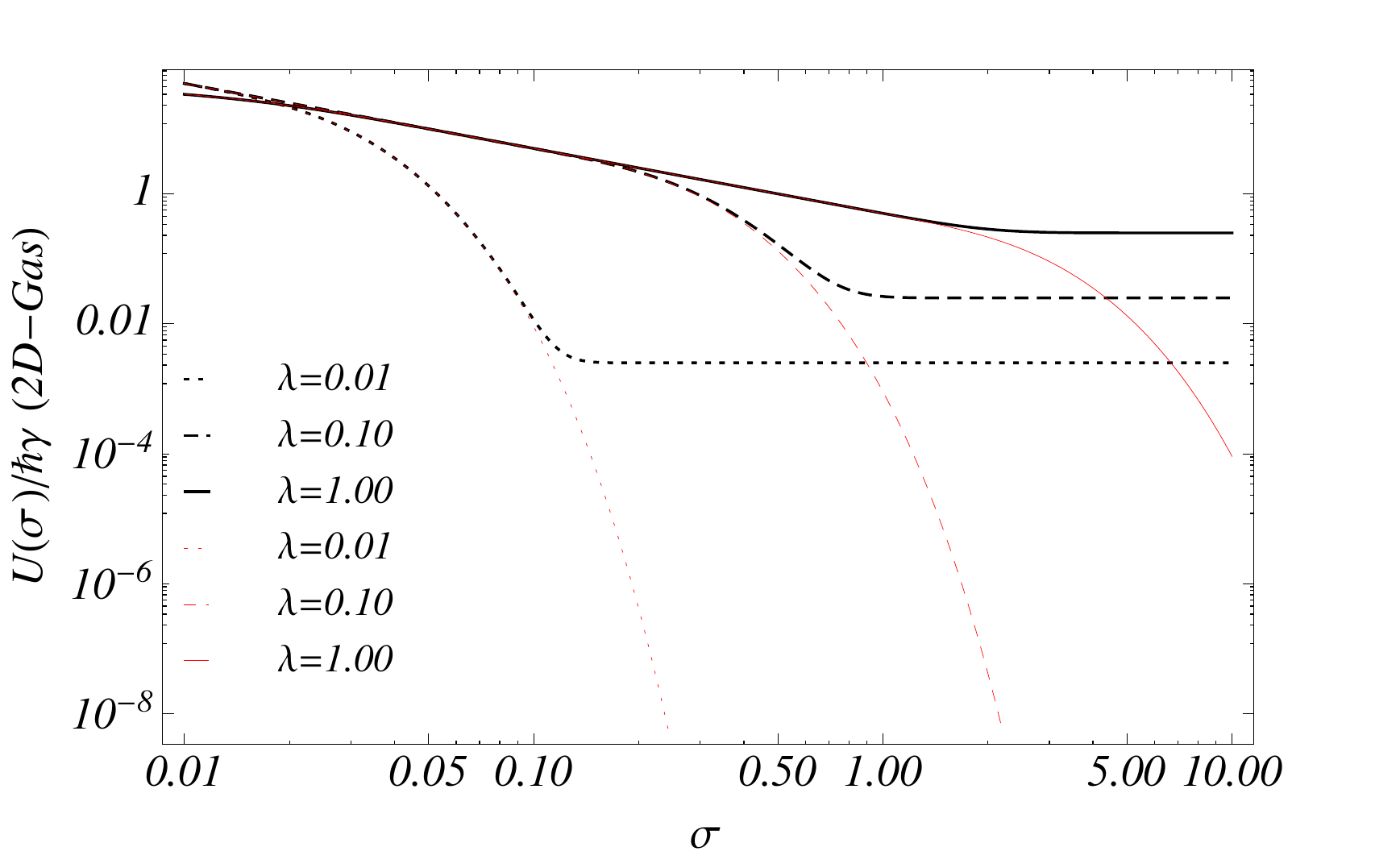}
\includegraphics[width= 8.5cm]{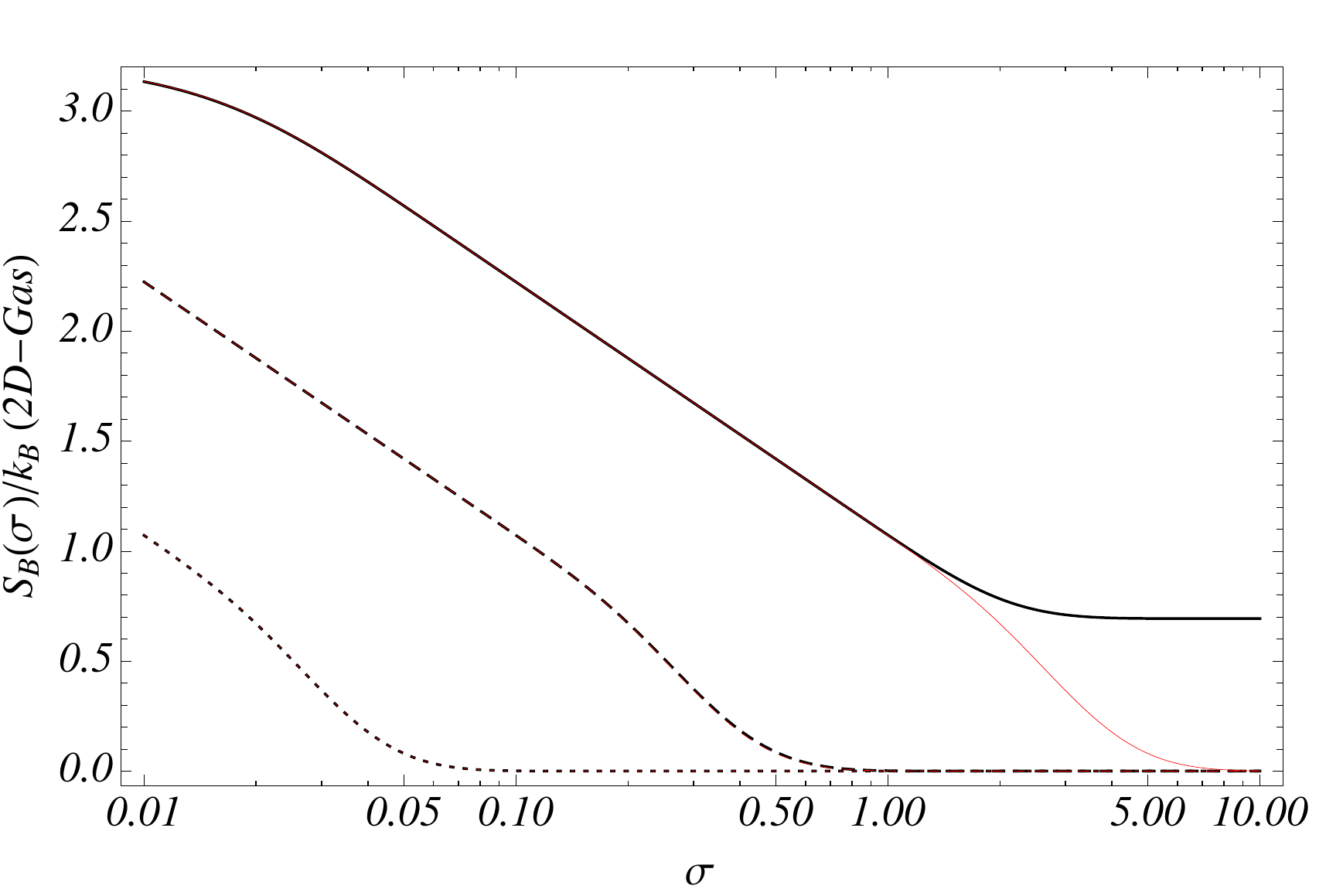}
\includegraphics[width= 8.5cm]{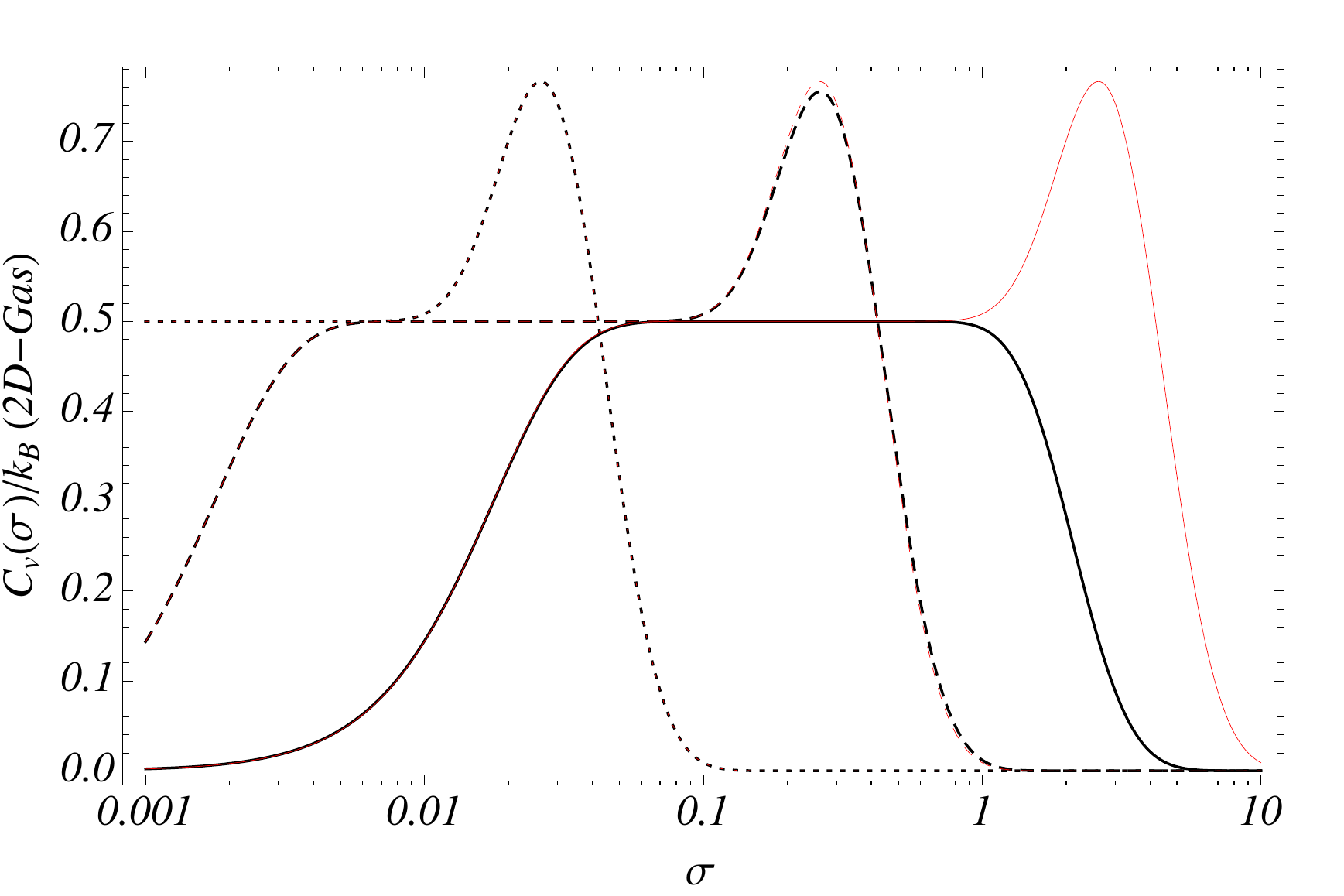}
\caption{Internal energy, $U(\sigma)$, Boltzmann entropy, $S_k(\sigma)$, and the heat capacity, $C_v(\sigma)$, of a gas of $2D$ quantum rotors as function of $\sigma$.
One has considered $\lambda  = 0.01$ (dotted lines), $0.1$ (dashed lines) and $1$ (solid lines) for NC (black lines) and standard (red lines) QM systems.}
\label{stFig.3}
\end{figure}

\begin{figure}
\includegraphics[width= 9.3cm]{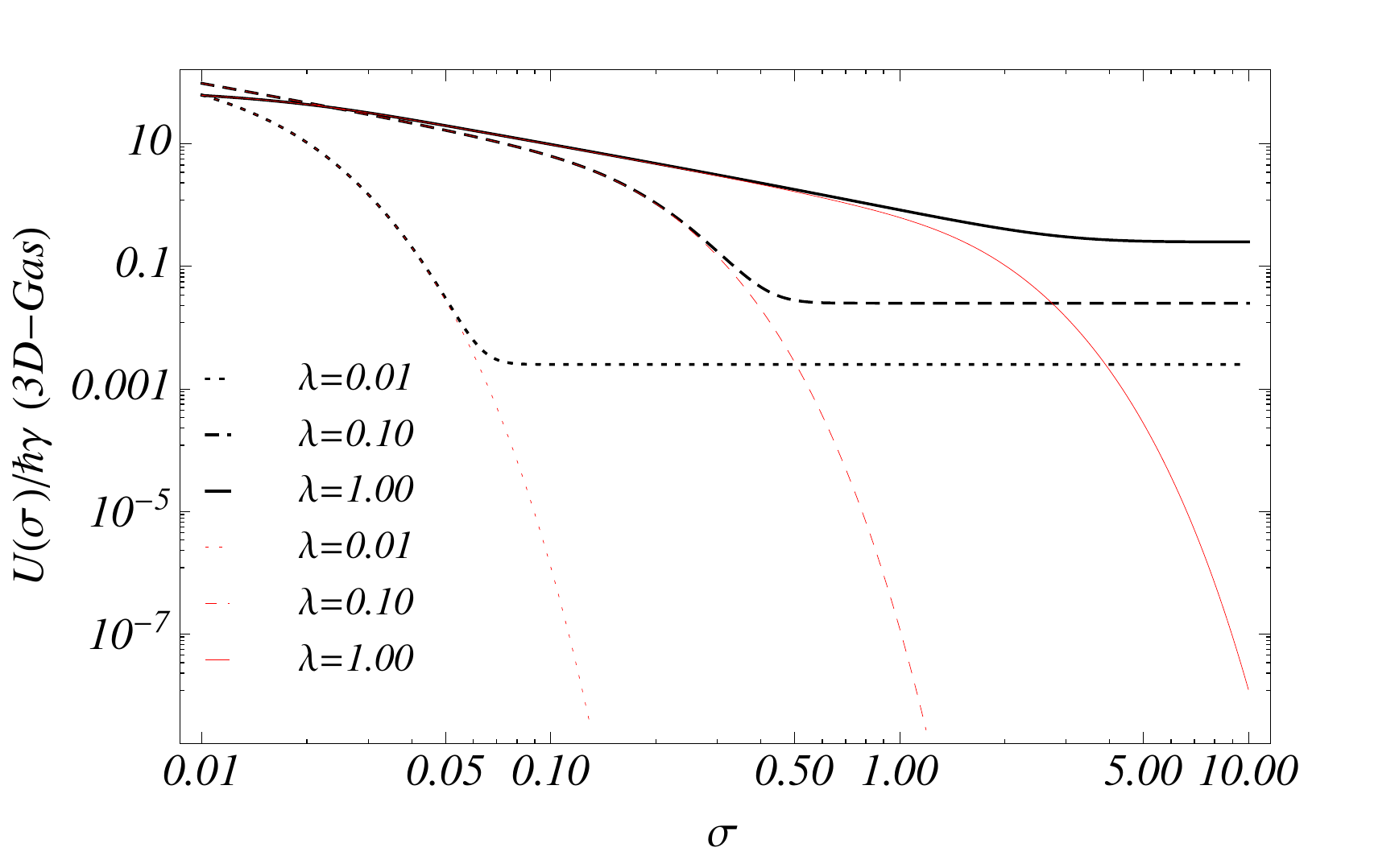}
\includegraphics[width= 8.5cm]{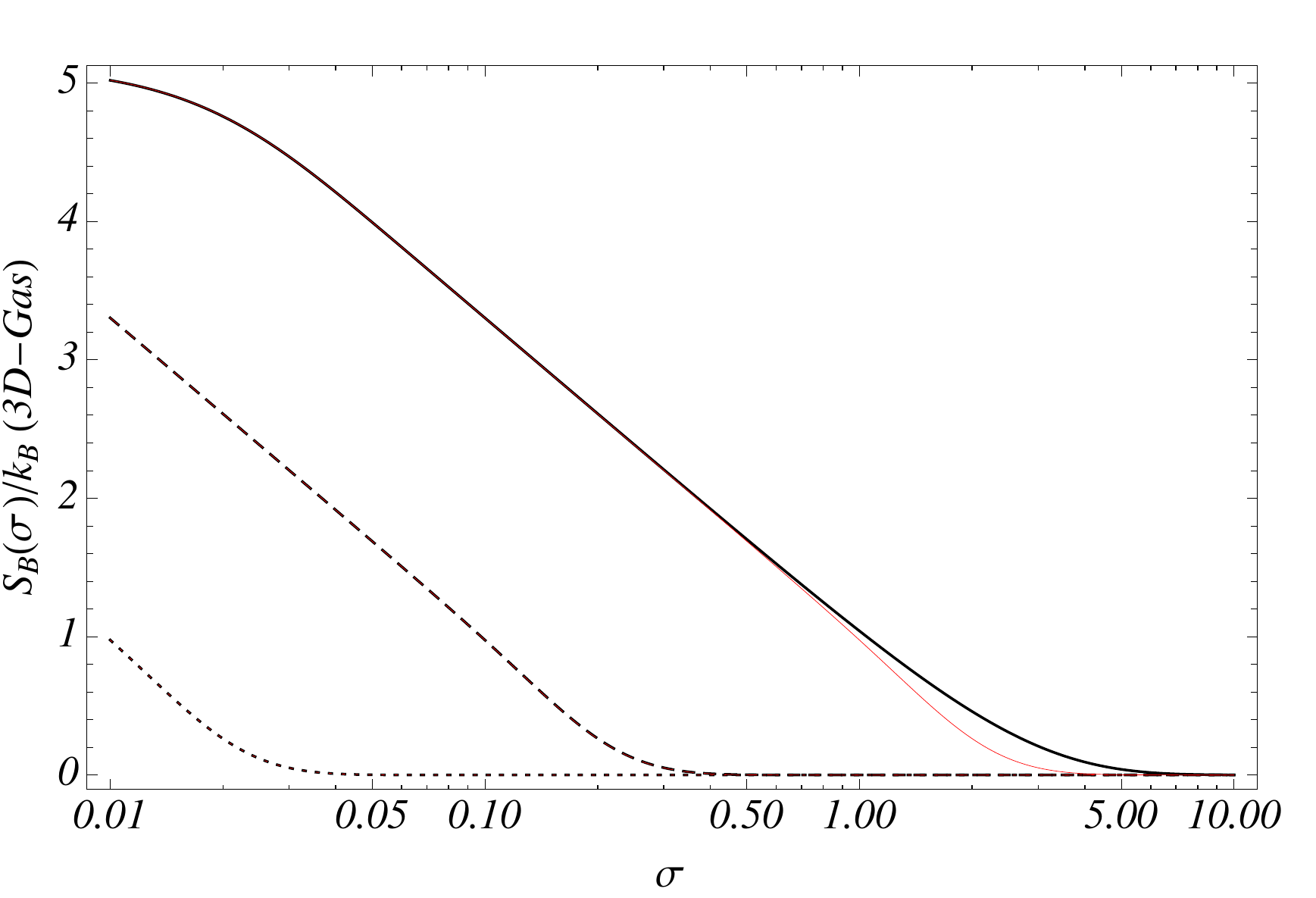}
\includegraphics[width= 8.5cm]{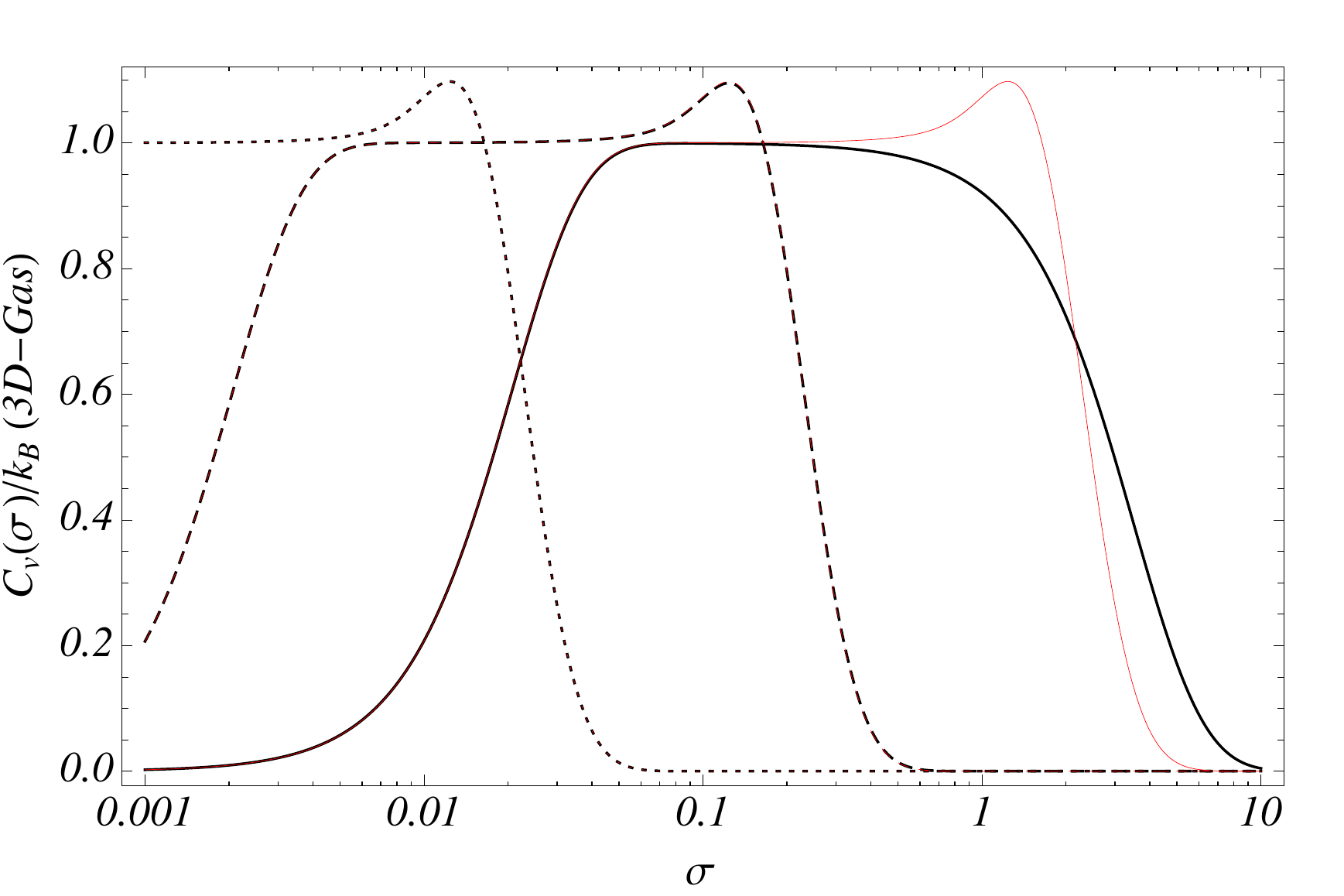}
\caption{Internal energy, $U(\sigma)$, Boltzmann entropy, $S_k(\sigma)$, and the heat capacity, $C_v(\sigma)$, of a gas of $3D$ quantum rotors as function of $\sigma$.
One has considered $\lambda  = 0.01$ (dotted lines), $0.1$ (dashed lines) and $1$ (solid lines) for NC (black lines) and standard (red lines) QM systems.}
\label{stFig.4}
\end{figure}

The quantitative effects for discrete values of $\lambda = 0.01,\, 0.1,$ and $1$ are depicted in 
Figs.~\ref{stFig.3} and \ref{stFig.4} for $2D$ and $3D$ quantum gases, respectively.
Again, NC results (solid black lines) are compared with the standard QM ones (dashed red lines).
In particular, one notices that the internal energy in the NC configuration exhibits well defined {\em plateaus} due to the NC quantization aspects. 
Once again, the NC effects are evinced for larger values of $\sigma$ and $\lambda$.
An additional typical quantum behavior is identified through the entropy of a gas of $2D$ NC quantum rotors: specifically for $\lambda = 1$ one finds $S_B \sim k_B \ln[2]$ as depicted in the second plot of Fig.~\ref{stFig.3}.

\begin{figure}
\includegraphics[width= 17cm]{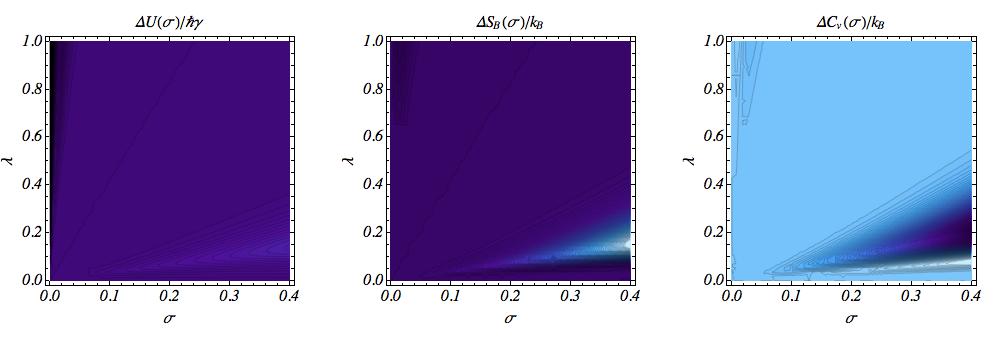}
\includegraphics[width= 17cm]{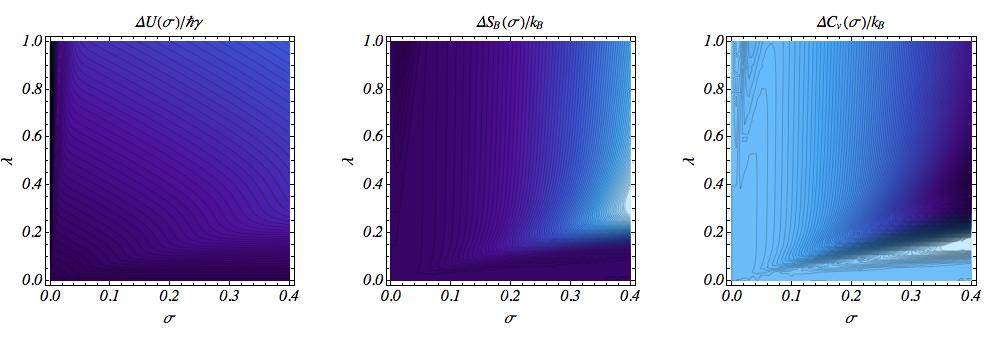}
\caption{Classical limit region ($\sigma < 1$) of NC deviations from standard results for internal energy, $U(\sigma)$, Boltzmann entropy, $S_k(\sigma)$, and the heat capacity, $C_v(\sigma)$, of gases of $2D$ (first row) and $3D$ (second row) quantum rotors as function of $\sigma$ and $\lambda$. 
The color scheme is reproduced from Fig.~\ref{stFig.1}.}
\label{stFig.5}
\end{figure}

Finally, one can consider the possibility of extending our results to the classical limit, i. e. lower values of $\sigma$ corresponding to higher temperatures, under which the NC and any kind of QM effect are used to be attenuated.
Fig.~\ref{stFig.5} roughly shows the classical limits of NC deviations from standard QM for $2D$ and $3D$ configurations.

\section{Conclusions}

The implications of noncommutativity on the quantum behavior related to decoherence and missing information issues of $2D$ and $3D$ free particle systems has been investigated, and the thermodynamic variables related to that and to the corresponding $2D$ and $3D$ quantum rotor systems has been obtained.
Quantization and thermodynamic aspects evincing the deviation from standard QM results have been discussed.

As noticed in some previous works \cite{Bernardini13A,Bernardini13B}, the phase-space noncommutativity and, in particular, the noncommutativity between momentum coordinates, induce the destruction of the original features of the wave function for the commutative problem and distort the stationary behavior of quantum systems by introducing missing information elements.
The modification of quantized energy levels is also reflected on the thermodynamic limit of such NC quantum systems.
In particular, some changes on the behavior of the internal energy, of the Boltzmann entropy and of the heat capacity derived from the partition function have been obtained for the thermalized $2D$ and $3D$ gases of free particles and quantum rotors.
The correspondence between the NC effects over quantum and classical variables were evidenced, and it has been shown that the most suitable range of temperatures for detecting noncommutativity in the phase-space, in particular, due to momentum coordinates, is around $\sim \eta/ 2 m k_B$ (c. f. Eq.~(\ref{steq37})).

All the results, even those related to the thermodynamic limit, can be completely re-interpreted in terms of an analogy with the Zeeman-effect, resulting in a description of the system with well-defined NC quantum numbers and a factorable wave-function.
For quantum rotors, the same coupling with a kind of magnetic field given by $B \sim \eta/\hbar q$ can be identified.
Our results suggest that phase-space noncomutativity effects can be interestingly considered when addressing the issues of interface between quantum and classical descriptions of Nature.

{\em Acknowledgments - The work of AEB is supported by the Brazilian Agencies FAPESP (grant 12/03561-0) and CNPq (grant 300809/2013-1). The work of CB is partially supported under the Portuguese Funda\c{c}\~ao para Ci\^encia e Tecnologia (FCT) under the grant SFRH/BPD/62861/2009.}
\pagebreak

\renewcommand{\baselinestretch}{1}

\end{document}